\documentclass[useAMS,usenatbib]{mn2e}
\usepackage{amsmath}
\usepackage{txfonts}
\usepackage{graphics}
\usepackage{graphicx}
\usepackage{times}
\usepackage{subfigure}
\usepackage{mn2e-breakabs}

\newcommand{\mnras}{MNRAS}

\newcommand{\physrep}{Physics Reports}

\newcommand{\Mpcoh}{\ h^{-1} {\rm Mpc}}

\begin{document}
\voffset-1.25cm
\title[Anisotropic clustering in CMASS galaxies]{The clustering of galaxies in
the SDSS-III Baryon Oscillation Spectroscopic Survey: measurements of the growth
of structure and expansion rate at z=0.57 from anisotropic clustering}
\author[Reid et al.]{
\parbox{\textwidth}{
Beth A. Reid$^{1,2}$\thanks{E-mail: beth.ann.reid@gmail.com}, Lado
Samushia$^{3,4}$, Martin White$^{1,5}$, Will J. Percival$^{3}$, Marc Manera$^{3}$,
Nikhil Padmanabhan$^{6}$, Ashley J. Ross$^{3}$, Ariel G. S\'anchez$^{7}$,  
Stephen Bailey$^{1}$, Dmitry
Bizyaev$^{8}$, Adam S.~Bolton$^{9}$, Howard
Brewington$^{8}$, J. Brinkmann$^{8}$, Joel R.~Brownstein$^{9}$,  Antonio J. Cuesta$^{6}$, Daniel J.
Eisenstein$^{10}$, James E.~Gunn$^{11}$, 
Klaus Honscheid$^{12}$, Elena Malanushenko$^{8}$, Viktor Malanushenko$^{8}$, Claudia
Maraston$^{3}$, Cameron K.~McBride$^{10}$, Demitri Muna$^{13}$, Robert C.~Nichol$^{3}$, Daniel
Oravetz$^{8}$,  Kaike Pan$^{8}$, Roland de
Putter$^{14,15}$, N.~A.~Roe$^{1}$, Nicholas P. Ross$^{1}$, David J.~Schlegel$^{1}$, Donald P.
Schneider$^{16}$, Hee-Jong Seo$^{17}$, Alaina Shelden$^{8}$, Erin S. Sheldon$^{18}$, Audrey
Simmons$^{8}$, Ramin A.~Skibba$^{19}$, Stephanie Snedden$^{8}$, Molly E.
C.~Swanson$^{10}$, Daniel Thomas$^{3}$, Jeremy Tinker$^{13}$, Rita Tojeiro$^{3}$,  
Licia Verde$^{15,20}$, David A. Wake$^{21}$, Benjamin A.
Weaver$^{13}$, David H. Weinberg$^{22}$, Idit Zehavi$^{23}$, 
Gong-Bo Zhao$^{3,24}$
}
\vspace*{4pt} \\
$^{1}$ Lawrence Berkeley National Laboratory, 1 Cyclotron Road, Berkeley, CA 94720, USA \\
$^{2}$ Hubble Fellow \\
$^{3}$ Institute of Cosmology and Gravitation, University of Portsmouth, Dennis Sciama Building, Portsmouth, P01 3FX, U.K.  \\
$^{4}$ National Abastumani Astrophysical Observatory, Ilia State University, 2A Kazbegi Ave., GE-1060 Tbilisi, Georgia  \\
$^{5}$ Departments of Physics and Astronomy, University of California, Berkeley, CA 94720, USA \\
$^{6}$ Department of Physics, Yale University, 260 Whitney Ave, New Haven, CT 06520, USA \\
$^{7}$ Max-Planck-Institut f\"{u}r extraterrestrische Physik, Postfach 1312,
Giessenbachstr., 85741 Garching, Germany \\
$^{8}$ Apache Point Observatory, P.O. Box 59, Sunspot, NM 88349-0059, USA \\
$^{9}$ Department of Physics and Astronomy, University of Utah, 115 S 1400 E,
Salt Lake City, UT 84112 USA \\
$^{10}$ Harvard-Smithsonian Center for Astrophysics, 60 Garden St., Cambridge, MA 02138, USA \\
$^{11}$ Department of Astrophysical Sciences, Princeton University, Princeton, NJ 08544, USA \\
$^{12}$ Department of Physics and CCAPP, Ohio State University, Columbus, OH, USA \\
$^{13}$ Center for Cosmology and Particle Physics, New York University, New York, NY 10003, USA \\
$^{14}$ Instituto de Fisica Corpuscular, Universidad de Valencia-CSIC, Spain \\
$^{15}$ ICC, University of Barcelona (IEEC-UB), Marti i Franques 1, Barcelona 08028, Spain \\
$^{16}$ Department of Astronomy and Astrophysics and Institute for Gravitation
and the Cosmos, The Pennsylvania State University, University Park, PA 16802, USA \\
$^{17}$ Berkeley Center for Cosmological Physics, LBL and Department of Physics, University of California, Berkeley, CA 94720, USA \\
$^{18}$ Brookhaven National Laboratory, Bldg 510, Upton, New York 11973, USA \\
$^{19}$ Steward Observatory, University of Arizona, AZ, USA \\
$^{20}$ ICREA \\
$^{21}$ Yale Center for Astronomy and Astrophysics, Yale University, New Haven,
CT, 06520, USA \\
$^{22}$ Department of Astronomy and CCAPP, Ohio State University, Columbus, OH,
USA \\
$^{23}$ Department of Astronomy, Case Western Reserve University, OH, USA \\
$^{24}$ National Astronomy Observatories, Chinese Academy of Science, Beijing, 100012, P.R.China
}

\date{\today} 
\pagerange{\pageref{firstpage}--\pageref{firstpage}}

\maketitle

\label{firstpage}

\begin{abstract}
We analyze the anisotropic clustering of massive galaxies from the 
Sloan Digital Sky Survey III Baryon 
Oscillation Spectroscopic Survey (BOSS) Data Release 9 (DR9) sample, which
consists of 264283 galaxies in the redshift range $0.43 < z < 0.7$ spanning 3275 
square degrees.  Both peculiar velocities and errors in the assumed
redshift-distance relation (``Alcock-Paczynski effect'') generate correlations
between clustering amplitude and orientation with respect to the line-of-sight.  
Together with the sharp baryon acoustic
oscillation (BAO) standard ruler, our measurements of the broadband shape of the monopole and
quadrupole correlation functions simultaneously constrain 
the comoving angular diameter distance ($2190 \pm 61$ Mpc) to $z=0.57$, the Hubble
expansion rate at $z=0.57$ ($92.4 \pm 4.5$ km s$^{-1}$ Mpc$^{-1}$), and the growth
rate of structure at that same redshift (${\rm d}\sigma_8/{\rm d}\ln a = 0.43 \pm 0.069$).  
Our analysis provides the best current direct determination of both $D_A$ and
$H$ in galaxy clustering data using this technique.
If we further assume a $\Lambda$CDM expansion history, our growth constraint
tightens to ${\rm d}\sigma_8/{\rm d} \ln a = 0.415 \pm 0.034$.  
In combination with the cosmic microwave background, our measurements of 
$D_A$, $H$, and ${\rm d}\sigma_8/{\rm d} \ln a$ all separately require dark energy at $z > 0.57$,
and when combined imply $\Omega_{\Lambda} = 0.74 \pm 0.016$, independent of
the Universe's evolution at $z<0.57$.
In our companion paper \citep{Samushia:prep}, 
we explore further cosmological implications of these observations.
\end{abstract}


\pagebreak
\section{Introduction}
Measurements of the cosmic distance-redshift relation using supernovae
\citep{riess:1998,perlmutter:1999,Kes09,amanullah:2010}, the cosmic microwave
background \citep{WMAP7}, the Hubble constant \citep{riess:2011}, and baryon
acoustic oscillations
\citep[BAO;][]{Eis05,Col05,Hut06,Pad07,BCBL,Per07,Oku08,Gaz09,Kaz10,Per10,Rei10,Bla11a,Beutler11}
have revealed that the expansion of the universe is accelerating; either the
energy density of the universe is dominated by an exotic ``dark energy'', or
general relativity requires modification. The observed anisotropic clustering of
galaxies can help distinguish between these possibilities by allowing
simultaneous measurements of both the geometry of the Universe and the growth rate
of structure.

Galaxy redshift surveys provide a powerful measurement of the growth rate through
redshift-space distortions (RSD) \citep{Kai87}. Although we expect the clustering
of galaxies in real space to have no preferred direction, galaxy maps produced
by estimating distances from redshifts obtained in spectroscopic surveys reveal
an anisotropic galaxy distribution \citep{ColFisWei95,Pea01,Per04,Ang08,Oku08,Guz08,SamPerRac11,Bla11b}.  
This anisotropy arises
because the recession velocities of galaxies, from which distances are inferred, include
components from both the Hubble flow and from peculiar velocities driven by the
clustering of matter (see \citealp{HamiltonReview} for a review).  Despite the
fact that galaxy light does not faithfully trace the mass, even on large scales,
galaxies are expected to act nearly as test particles within the cosmological
matter flow.  Thus the motions of galaxies carry an imprint of the rate of
growth of large-scale structure and allow us to both probe dark energy and test
General Relativity \cite[see e.g.][for recent
studies]{Jain08,NesPer08,Song09a,Song09b,PerWhi08,McDSel09,WhiSonPer09,Song10,Zhao10,Song11}.

The observed BAO feature in the power-spectrum and correlation function of galaxies
has been used to provide strong constraints on the geometry of the
Universe. While the BAO method is expected to be highly robust to systematic
uncertainties \citep[see, e.g.,][]{EisWhi04,PadWhi09,Seo10,Mehta11}, it does
not exploit the full information about the cosmological model available in
the two-dimensional clustering of galaxies.  Additional geometric information is
available by comparing the amplitude of clustering along and perpendicular to
the line-of-sight (LOS); this is known as the Alcock-Paczynski (AP) test
\citep{AP,Bal96}.  
RSD and AP tests rely on the measurements of anisotropy in the statistical
properties of the galaxy distribution and are partially degenerate with each other, 
so that constraints on the growth of structure from
RSD depend on the assumptions about the background geometry and vice versa 
\citep{Samushia11}.  
However, given high precision clustering measurements over a wide range of
scales, this degeneracy can be broken since RSD and AP have different
scale-dependences.  
Recently, the WiggleZ survey \citep{Drinkwater10} has performed 
a joint RSD and AP analysis that constrains the expansion history in 4 bins
across $0.1 < z < 0.9$ at the 10-15 per cent level \citep{Bla11c}.  Using the SDSS-II
LRG sample, \citet{ChuWang11} perform a similar analysis to the present work.
They measure the angular diameter distance $D_A(z=0.35) = 1048^{+60}_{-58}$ Mpc and 
the Hubble expansion rate $H(z=0.35) = 82.1^{+4.8}_{-4.9}$ 
km s$^{-1}$ Mpc$^{-1}$ at $z=0.35$ after marginalizing over redshift space distortions and
other parameter uncertainties. 

Obtaining reliable cosmological constraints from the RSD and AP effects
demands precise modeling of the nonlinear evolution of both the matter density and
velocity fields, as well as the ways in which the observed galaxies trace those
fields.  The halo model of large-scale
structure and variants thereof assume that galaxies form and evolve in the
potential wells of dark matter halos, and provides a successful means of
parametrizing the relation between the galaxy and halo density and velocity
fields.  Our modeling approach, based on \citet{ReiWhi11},
 uses perturbation theory to account for the
nonlinear redshift space clustering of halos in the quasilinear regime as a
function of cosmological parameters, and then uses the halo model framework to
motivate our choice of nuisance parameters describing the galaxy-halo relation.
We test these assumptions with a large volume of mock galaxy catalogs derived
from $N$-body simulations.

The ongoing Baryon Oscillation Spectroscopic Survey \citep[BOSS;][]{BOSS}, which
is part of Sloan Digital Sky Survey III (SDSS-III) \citep{Eis11}, 
is measuring spectroscopic redshifts of
1.5 million galaxies, approximately volume limited to $z\simeq 0.6$ (in addition
to spectra of 150,000 quasars and various ancillary observations).  The galaxies
are selected from the multi-color SDSS imaging to probe large-scale structure at
intermediate redshift; they trace a large cosmological volume while having
high enough number density to ensure that shot-noise is not a dominant
contributor to the clustering variance \citep{Whi11}.  The resulting clustering
measurements provide strong constraints on the parameters of standard
cosmological models.

We use the CMASS sample of BOSS galaxies that will be included in
SDSS Data Release 9 (DR9) to constrain the growth of structure and geometry
of the Universe. We apply RSD and AP tests to the data to measure the growth
rate, the Hubble expansion rate, and the comoving angular diameter distance
at $z_{\rm eff}=0.57$.
We improve on previous, similar analyses in a number of ways.  
First, we use a model for the nonlinear anisotropic correlation function that
is accurate to well below our statistical errors over the wide range of scales
between $25\Mpcoh$ and $160\Mpcoh$, which we validate using
68\,($h^{-1}$Gpc)$^{3}$ of $N$-body simulations populated with mock CMASS
galaxies.
Next, rather than assuming a fixed underlying linear matter power spectrum,
we use a prior on $P(k)$ based on the WMAP7 cosmic microwave background
constraints \citep{WMAP7,WMAP7cosmo} and marginalize over the remaining
uncertainties for all fits.  In addition to joint constraints on the geometric
(BAO and AP) and peculiar velocity (RSD) parameters, we present three ``null''
tests of the $\Lambda$CDM model.  
To begin, we simply ask whether any points in the $\Lambda$CDM parameter 
space allowed by WMAP7 provide a good fit to the CMASS clustering; in this case,
the only free parameters are those describing how galaxies trace matter.  In the
other two tests, we fit for the amplitude of peculiar velocities using
WMAP7 priors on geometric quantities, or we fit for the geometric parameters
with WMAP7 priors on RSD.  
Thus we can present the statistical precision with which our data measure
either peculiar velocities or the AP effect in the $\Lambda$CDM model.
Finally, given the strong detection of the BAO feature in the monopole of
the correlation function, we can break the degeneracy between
$(1+z_{\rm eff})D_A$ and $H$ with our AP measurement.
We present the most constraining measurement of $H(z_{\rm eff})$ from galaxy
clustering data to date using this technique, even after marginalizing over the
amplitude of the RSD effect.  
As our constraints exploit the full shape of the monopole and quadrupole
correlation functions, they rely on further assumptions about the cosmological
model: Gaussian, adiabatic, power-law primordial perturbations, the standard
number $N_{\rm eff}=3.04$ of massless neutrino species
\citep[see discussion in][]{WMAP7cosmo}, and that dark matter is ``cold''
on the relatively large scales of interest.

This paper is organized as follows. In Sec.~\ref{sec:data} we describe BOSS DR9
CMASS data and in Sec.~\ref{sec:measurements} we describe the measurements of
two-dimensional clustering of galaxies used in this analysis. Sec.~\ref{sec:theory} 
reviews the theory of the RSD and AP effects, and describes 
the theoretical model used to fit our measurements.
Sec.~\ref{sec:analysis} outlines the methods we use to analyse
the data and Sec.~\ref{sec:results} presents the results of analysis. We
conclude by discussing the cosmological implications of our results in 
Sec.~\ref{sec:discussion}.

\section{Data}
\label{sec:data}
\begin{figure}
  \includegraphics[width=85mm]{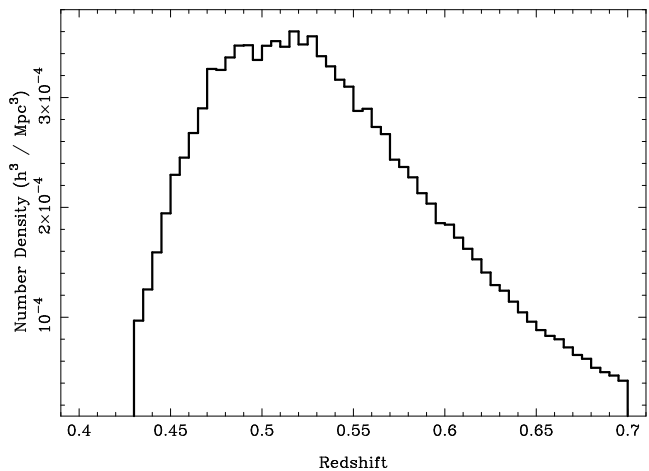}
  \caption{Number density as a function of redshift for the CMASS
  galaxies used in this analysis.  After accounting for our weighting scheme,
the effective redshift for galaxy pairs in this sample is $z_{\rm eff} = 0.57$.}
  \label{fig:Nz}
\end{figure}

BOSS targets for spectroscopy luminous galaxies selected from the multi-color
SDSS imaging \citep{Fuk96,Gun98,Yor00,Gun06}.
The target selection algorithms are summarized by \citet{Eis11} and
\citet{Aardvark}.
For the galaxy sample referred to as ``CMASS'', color-magnitude cuts are
applied to select a roughly volume-limited sample of
massive, luminous galaxies; see \citet{Masters11} for a detailed examination of
the properties of CMASS targets in the COSMOS field.
The majority of the galaxies are old stellar systems whose prominent
4000$\,\AA$ break in their spectral energy distributions makes them relatively
easy to select in multi-color data. Most CMASS galaxies are
central galaxies residing in dark matter halos of $10^{13}\,h^{-1}M_\odot$, but
a non-negligible fraction are satellites which live primarily in halos about 10
times more massive \citep{Whi11}. These galaxies are intrinsically very luminous
and at the high mass end of the stellar mass function \citep{Mar11}.
Galaxies in the CMASS sample are highly biased \citep[$b\sim 2$,][]{Whi11}.
In addition, they trace a
large cosmological volume while having high enough number density to ensure that
shot-noise is not a dominant contributor to the clustering variance, which makes
them particularly powerful for probing statistical properties of large-scale
structure.  
 
\begin{figure}
  \includegraphics[width=85mm]{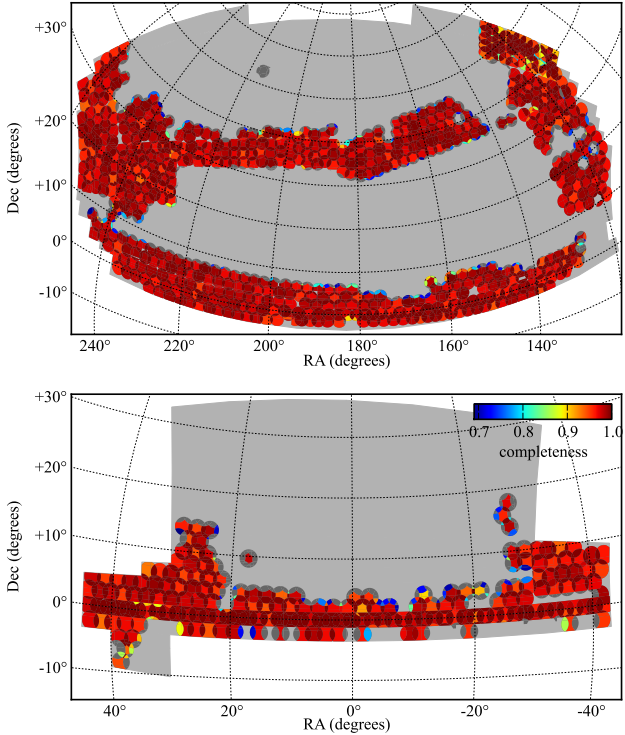}
  \caption{BOSS DR9 sky coverage.  The light gray region shows the expected
total footprint of the survey, while the colors indicate completeness in the DR9
spectroscopy for regions included in our analysis.  
Dark grey regions are removed from the analysis by completeness
or redshift failure cuts; see Sec.~3.5 of \citet{Aardvark} for further details.}
  \label{fig:mask}
\end{figure}

\citet{Aardvark} details the steps for generating the large-scale structure
catalog and mask for DR9, which includes the data taken by
BOSS through July 2011 and covers 3275 $\rm deg^2$ of sky.
In our analysis we use galaxies from the BOSS CMASS DR9 catalog in the
redshift range of $0.43<z<0.70$. The sample includes a total of 264,283 galaxies,
207,246 in the north and 57,037 in the south Galactic hemispheres.
Figure~\ref{fig:Nz} shows the redshift distribution and Figure~\ref{fig:mask}
shows the angular distribution of the galaxies in our sample.

\section{Measurements}
\label{sec:measurements}
\subsection{Two point statistics}
\label{sec:meas}
To compute redshift space separations for each pair of galaxies given their
angular coordinates and redshifts, we must adopt a cosmological model.  We use
the same one as used to generate our mock catalogs, namely spatially-flat
$\Lambda$CDM with $\Omega_m=0.274$.
This is the same cosmology as assumed in \citet{Whi11} and in our companion
papers \citep{Aardvark,Manera:2012,Ross:2012,Sanchez:2012,Samushia:prep,Tojeiro:2012}.
Our model accounts for this assumption and scales the theory prediction
accordingly when testing a cosmological model with a different
distance-redshift relation; see Sec.~\ref{APsec}.

Using the galaxy catalog of \citet{Aardvark} we compute weighted
``data-data'' (DD) pair counts in bins of $s$ and $\mu$ 
\begin{equation}
  DD(s_i, \mu_j) = \displaystyle\sum_{k=1}^{N_{\rm tot}}\displaystyle\sum_{l=k+1}^{
  N_{\rm tot}} \Theta_{k,l}(s_i,\mu_j) w_k(s_i)w_l(s_i),
  \label{eq:dd}
\end{equation}
\noindent
where $s$ is the comoving pair separation in redshift space, $\mu$ is the
cosine of the angle between the pair separation vector and the LOS and $w_k$
is the weight of $k^{\rm th}$ galaxy in the catalog.  The double sum
runs over all galaxies and $\Theta_{k,l}(s_i,\mu_j)=1$ if the pair separation between
two galaxies falls into bin $s_i, \mu_j$, and is zero otherwise. 

Three distinct effects contribute
to the final weight $w_i$ of each galaxy.  These weights are described in more
detail in \citet{Aardvark} and \citet{Ross:2012}.  
First, galaxies lacking a redshift
due to fiber collisions or because their spectrum was not adequate to secure a
redshift are accounted for by upweighting the
nearest galaxy by weight $w = (1+n)$, where $n$ is the number of near
neighbors without a redshift.   Second, we use the minimum variance
weighting of \citet{Ham93},
\begin{equation}
  w(s) = \frac{1}{1+J_3(s)\bar{n}(z)},
\end{equation}
where $\bar{n}(z)$ is the expected number density of galaxies at the redshift
of the galaxy and
\begin{equation}
  J_3(s) = 2\pi\displaystyle\int_0^s s'^2ds'd\mu\ \xi(s',\mu)
\end{equation}
is the angularly averaged redshift-space correlation function integrated up to
the separation of galaxies in the pair. For every galaxy this weight will vary
depending on which pair-counting bin it is assigned. For a constant radial
selection function this weighting scheme results in the minimal variance of the
estimated correlation function \citep[for details see][]{Ham93}.
Note that the analyses in \citet{Sanchez:2012} and \citet{Aardvark} use
 scale-independent weights; differences between the approaches are small in
practice.

The third weight corrects for angular systematics, related to the angular
variations in density of stars that make detection of galaxies harder in areas
of sky closer to the Galactic equator \citep[for details see][]{Ross:2012}.
The total weight is the product of these three weights.
We bin $s$ in 23 equal logarithmic bins between $s_{\rm min} = 25.1$ and $s_{\rm max} = 160\Mpcoh$
with dlog$_{10} s = 0.035$, and 200 equally-spaced $\mu$ bins between 0 and 1.
We compute ``data-random'' (DR) and
``random-random'' (RR) pair counts as in Eqn.~\ref{eq:dd}, except that each
random point is assigned only the $J_3(s)$ weight and not the close-pair
correction and systematic weights.
Positions of objects in our random catalog are generated using the
observational mask and redshifts are generated by picking a random redshift
drawn from the measured redshifts of observed galaxies.
Our random catalogs contain approximately 70 times more objects than the
galaxy catalog. 

\begin{figure*}
\begin{center}
\resizebox{82mm}{!}{\includegraphics{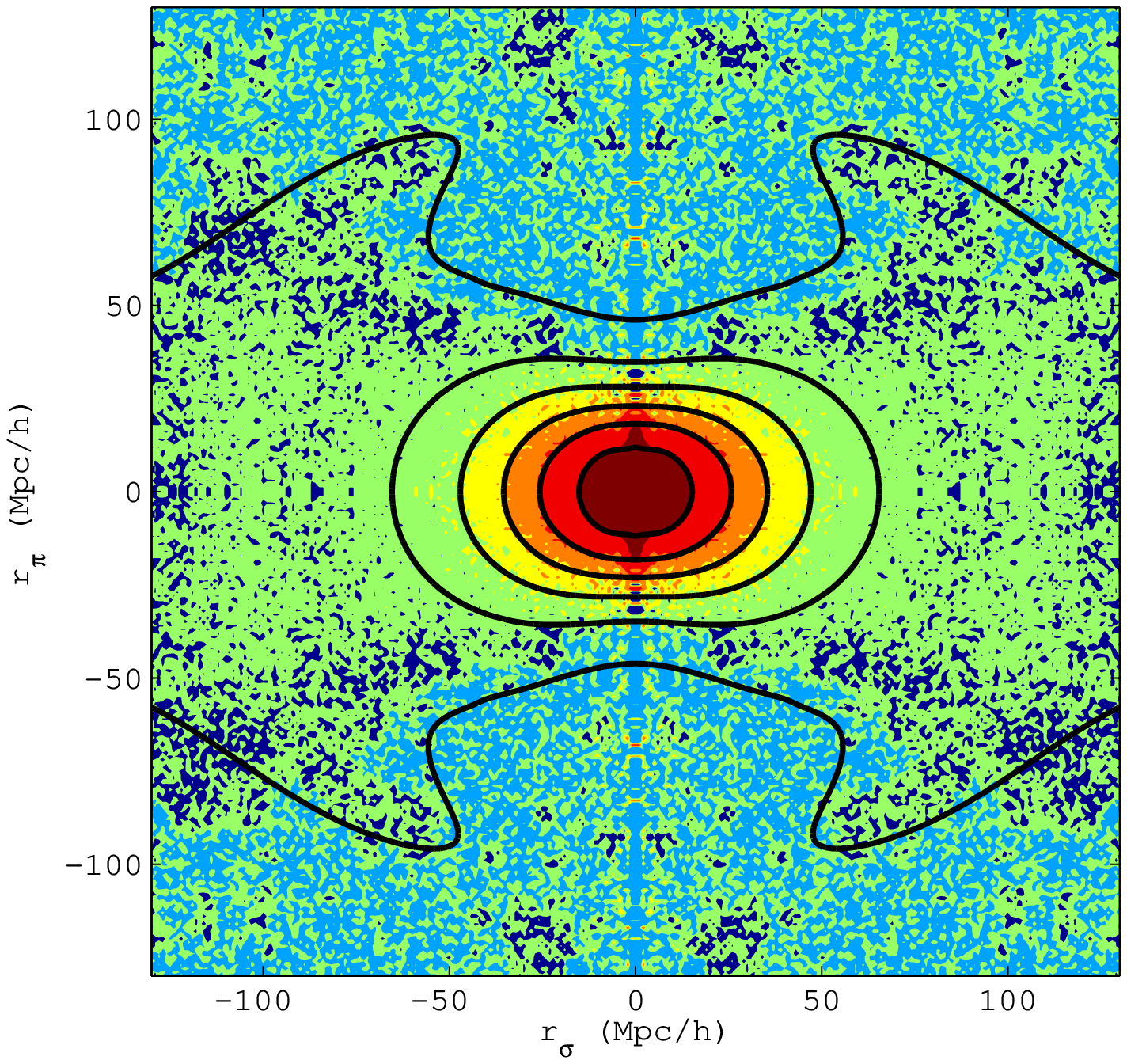}}
\resizebox{91mm}{!}{\includegraphics{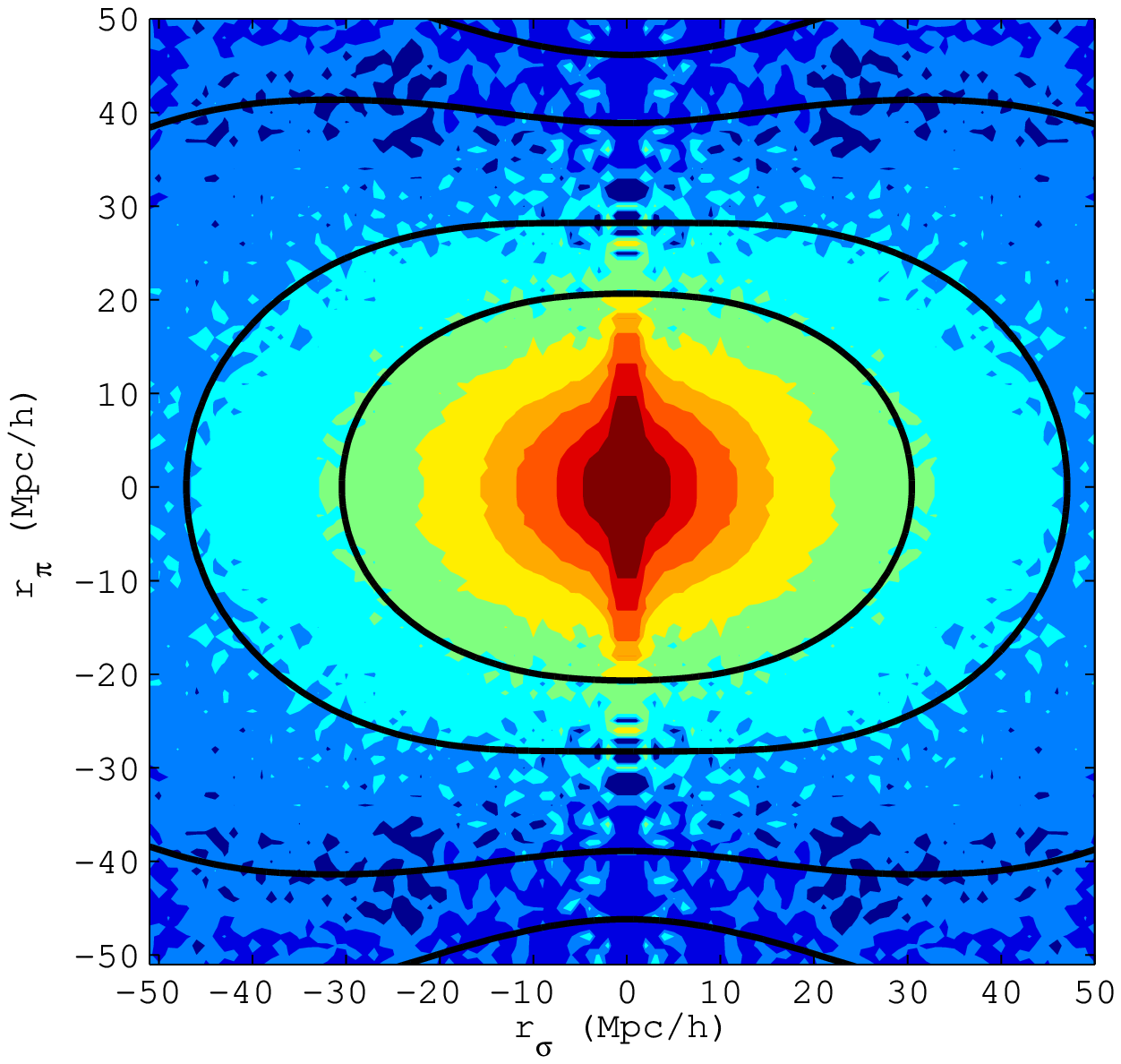}}
\caption{{\em Left panel}: Two-dimensional correlation function of CMASS galaxies (color)
compared with the best fit model described in Section \ref{sec:goodfitLCDM}
(black lines). Contours of equal $\xi$ are shown at [0.6, 0.2, 0.1, 0.05, 0.02,
0].  {\em Right panel}: Smaller-scale two-dimensional clustering.  We show
model contours at [0.14, 0.05, 0.01, 0].  The value of $\xi_0$ at the minimum
separation bin in our analysis is shown as the innermost contour. 
The $\mu \approx 1$ 
``finger-of-god'' effects are small on the scales we use in this analysis.}
\label{fig:butterfly}
\end{center}
\end{figure*}

Following \citet{LanSza93}, the pair counts are combined to estimate the
anisotropic correlation function as:
\begin{equation}
\hat{\xi}_{LS}(s_i, \mu_j) = \frac{DD(s_i, \mu_j) - 2DR(s_i, \mu_j) + RR(s_i,
  \mu_j)}{RR(s_i,\mu_j)}.
\end{equation}
Figure \ref{fig:butterfly} shows our measurement of $\hat{\xi}_{LS}(s_i, \mu_j)$
in terms of LOS separation $r_{\pi} = s\mu_s$ and transverse separation
$r_{\sigma} = s(1-\mu^2)^{1/2}$.  The central ``squashing'' is due to peculiar
velocities.  In the left panel, the BAO ridge at $\sim 100$ $h^{-1}$
Mpc is evident.  In the right panel, we show the clustering signal on smaller
scales; the ``finger-of-God'' effect is visible for small transverse
separations but small on the scales we analyse.  
The innermost contour in the right panel indicates the value of $\xi_0$ in the
smallest separation bin included in our cosmological analysis.

Rather than work with the two-dimensional correlation function
$\xi(s,\mu_s)$, we conduct our cosmological analysis on the first two
even Legendre polynomial moments, $\xi_0(s)$ and $\xi_2(s)$, 
defined by
\begin{equation}
\xi_{\ell}(s) = \frac{2\ell+1}{2}\int d\mu_s\ \xi(s, \mu_s) L_{\ell}(\mu_s),
\label{eq:moments}
\end{equation} 
or equivalently,
\begin{equation}
\xi(s, \mu_s) \equiv \sum_{\ell=0}^{\infty} \xi_{\ell}(s) L_{\ell}(\mu_s).
\end{equation} 
Here $L_\ell$ is the Legendre polynomial of order $\ell$.
By symmetry all odd-$\ell$ moments vanish and on large scales the measurements
become increasingly noisy to larger $\ell$.
The correlation functions $\hat{\xi}_0(s)$ and $\hat{\xi}_2(s)$ are
estimated from $\hat{\xi}_{LS}(s_i, \mu_j)$ using a Riemann sum to approximate
Eqn.~\ref{eq:moments}.  
We include all galaxy pairs between 25 and $160\,h^{-1}$Mpc in our analysis.
We also caution the reader that we have adopted logarithmically-spaced bins,
while our companion papers \citep{Aardvark,Ross:2012,Sanchez:2012} analyze 
clustering in linearly-spaced bins of differing bin sizes.
Our measurements of $\xi_0$ and $\xi_2$, along with diagonal errors estimated
from mock catalogs (\citealp{Manera:2012}; see Sec.~\ref{sec:covmat}) are shown
in Figure~\ref{fig:xilmeas}.
The effective redshift of weighted pairs of galaxies in our sample is $z=0.57$,
with negligible scale dependence for the range of interest in this paper.
For the purposes of constraining cosmological models, we will interpret our
measurements as being at $z=0.57$.

\subsection{Covariance Matrices}
\label{sec:covmat}
\begin{figure}
\begin{center}
\resizebox{85mm}{!}{\includegraphics{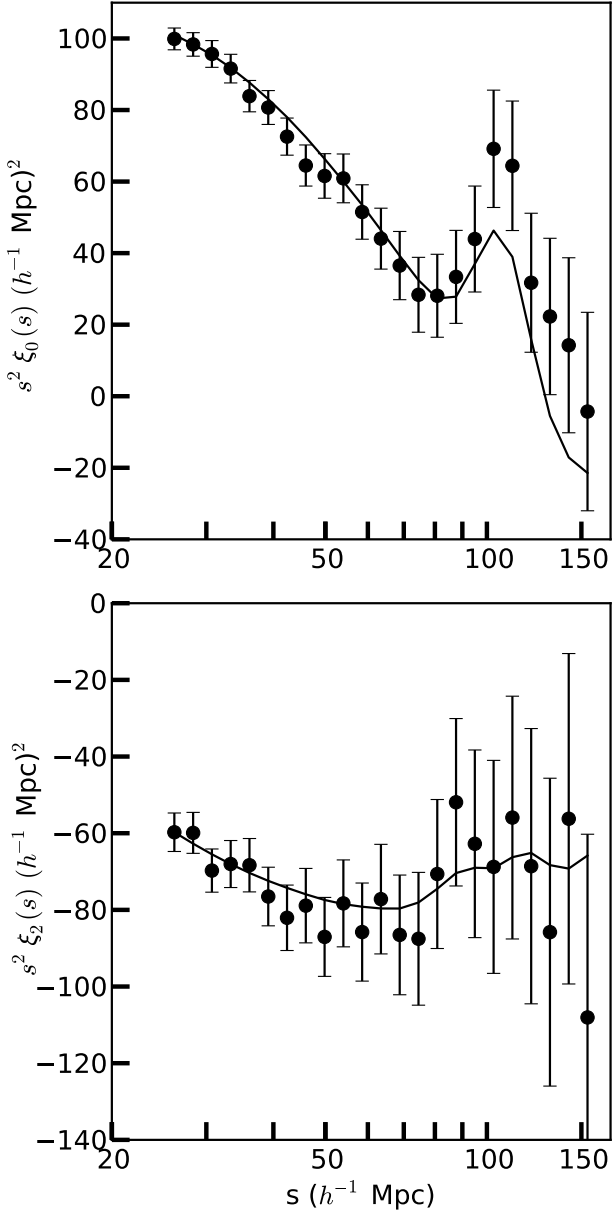}}
\end{center}
\caption{$\xi_0(s)$ and $\xi_2(s)$ measured from BOSS CMASS galaxies. The
errorbars correspond to diagonal elements of the covariance matrix.  The
best-fitting model described in Section \ref{sec:goodfitLCDM} is shown 
as the solid curve.  We use 23 logarithmically spaced bins and include pairs
between 25 and 160 $h^{-1}$ Mpc.}
\label{fig:xilmeas}
\end{figure}

The matrix describing the expected covariance of our measurements of
$\xi_\ell(s)$ in bins of redshift space separation depends in linear theory
only on the underlying linear matter power spectrum, the bias of the galaxies,
the shot-noise (often assumed Poisson) and the geometry of the survey.
We use 600 mock galaxy catalogs, based on Lagrangian perturbation theory
(LPT) and described in detail in \citet{Manera:2012}, to estimate the
covariance matrix of our measurements.
We compute $\xi_\ell(s_i)$ for each mock in exactly the same way as from the
data (Sec.~\ref{sec:meas}) and estimate the covariance matrix as
\begin{equation}
  \label{eq:cov}
  C^{\ell_1 \ell_2}_{ij} = 
  \frac{1}{599}\displaystyle\sum_{k = 1}^{600}
  \left(\xi_{\ell_1}^k(s_i) -
  \bar{\xi}_{\ell_1}(s_i)\right)\left(\xi_{\ell_2}^k(s_j) - 
  \bar{\xi}_{\ell_2}(s_j)\right),
\end{equation}
\noindent
where $\xi_\ell^k(s_i)$ is the monopole ($\ell = 0$) or quadrupole ($\ell = 2$) 
correlation function for pairs in the $i^{th}$ separation bin in the $k^{th}$ mock. 
$\bar{\xi}_\ell(s)$ is the mean value over all 600 mocks.
The shape and amplitude of the average two-dimensional correlation function computed from the
mocks is a good match to the measured correlation function of the CMASS galaxies; 
see \citet{Manera:2012} and \citet{Ross:2012} for more detailed comparisons.  
The square roots of the diagonal elements of our covariance matrix are shown as 
the errorbars accompanying our measurements in Fig.~\ref{fig:xilmeas}.  We will examine 
the off-diagonal terms in the covariance matrix via the correlation matrix, 
or ``reduced covariance matrix'', defined as 
\begin{equation}
  C^{\ell_1\ell_2, \rm red}_{ij} =
  C^{\ell_1\ell_2}_{ij}/\sqrt{C^{\ell_1\ell_1}_{ii} C^{\ell_2 \ell_2}_{jj}},
\label{eq:redcov}
\end{equation}
\noindent
where the division sign denotes a term by term division.

\begin{figure}
\begin{center}
\includegraphics[width=85mm]{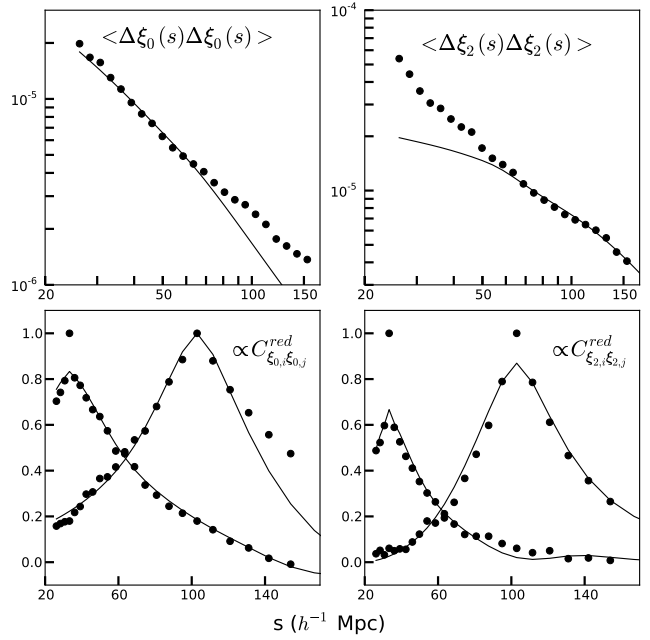}
\end{center}
\caption{{\em Upper panels}: Diagonal elements of the monopole and quadrupole
components of the covariance matrix
computed from the LPT-based mocks (points) compared with the linear theory
prediction (solid lines).  {\em Lower panels}: Two slices
through the reduced
covariance matrices $C^{00,{\rm red}}_{ij}$ and $C^{22,{\rm red}}_{ij}$ for 
separation bins of 33 and 103 $h^{-1}$ Mpc.  Linear theory predictions for the
reduced covariance matrices in the lower panels have been 
scaled by a constant factor to produce good agreement between linear theory and
mock covariances for off-diagonal elements, demonstrating that the scale
dependence of the off-diagonal terms matches the mock covariance matrix well,
but that there is extra diagonal covariance in the mocks compared with linear
theory. Elements of $C^{02}$ are small (not shown).}
\label{fig:LPTvslinear}
\end{figure}

In Figure \ref{fig:LPTvslinear} we compare selected slices of our mock
covariance matrix (points) to a simplified prediction from linear theory (solid
lines) that assumes a constant number density $\bar{n} = 3 \times 10^{-4}$ 
($h^{-1}$ Mpc)$^{-3}$ and neglects the effects of survey geometry \citep[see,
e.g.,][]{Teg97}.
\citet{Xu12} performed a detailed comparison of linear theory predictions with
measurements from the Las Damas SDSS-II LRG mock catalogs
\citep{McBride:prep}, and showed that a modified version of the linear theory
covariance with a few extra parameters provides a good description of the
$N$-body based covariances for $\xi_0(s)$.
The same seems to be true here as well.
The mock catalogs show a deviation from the naive linear theory
prediction for $\xi_2(s)$ on small scales; a direct consequence is that our
errors on quantities dependent on the quadrupole 
are larger than a simple Fisher analysis would indicate.  
We verify that the same qualitative behavior is seen for the diagonal elements of
the quadrupole covariance matrix in our smaller set of $N$-body
simulations used to calibrate the model correlation function.  This comparison
suggests that the LPT-based mocks are not underestimating the errors on $\xi_2$,
though more $N$-body simulations (and an accounting of survey geometry) 
would be required for a detailed check of the LPT-based mocks.

The lower panels of Figure \ref{fig:LPTvslinear} compare the reduced covariance
matrix to linear theory, where we have scaled the $C^{\rm red}_{ij}$
prediction from linear theory down by a constant, $c_i$.  This comparison
demonstrates that the scale dependences of the off-diagonal terms in the
covariance matrix are described well by linear theory, but that the nonlinear
evolution captured by the LPT mocks can be parametrized 
simply as an additional diagonal term.
Finally, while not shown here, the reduced covariances between $\xi_0$ and
$\xi_2$ are small.

Our analysis uses the LPT mock-based covariance matrix, which accurately 
accounts for both complexities of
the survey geometry as well as nonlinear corrections to the growth of structure
on the relatively large scales of interest here, and this allows us to
accurately report uncertainties associated with both our measurements and
parameter fits.

\section{Theory}
\label{sec:theory}

\subsection{Redshift Space Distortions: Linear Theory}

The effects of redshift space distortions in the linear regime are well-known
\citep[][see also  \citealp{HamiltonReview} for a comprehensive
review]{Kai87,Fis95}.  We briefly summarize them here.  The redshift-space
position, {\bf s}, of a galaxy differs from its real-space position, {\bf x},
due to its peculiar velocity,
\begin{equation} \label{eq:sx}
  {\bf s} = {\bf x} + v_z({\bf x})\,\widehat{\bf z},
\end{equation}
where $v_z({\bf x}) \equiv u_z({\bf x})/(aH)$ is the change in the apparent
LOS position of a galaxy due to the contribution of the LOS peculiar velocity $u_z$
to the galaxy's redshift.  
Since overdensities on large, linear scales grow in a converging velocity
field ($\nabla\cdot{\bf v}=-f\delta_m$), the effect of peculiar velocities
induces a coherent distortion in the measured clustering of galaxies that
allows us to measure the amplitude of the peculiar velocity field.
In linear theory, and with some approximations, the anisotropic galaxy power
spectrum becomes \citep{Kai87}
\begin{equation}
\label{eq:kaiser}
 P_g^s(k,\mu_k) = \left(b+f\mu_k^2\right)^2 P_m^r(k)
               = b^2\left(1+\beta\mu_k^2\right)^2 P_m^r(k)
\end{equation}
where $b$ is the linear galaxy bias, $\delta_g = b \delta_m$, $f\equiv d\ln
\sigma_8/d\ln a$ is the logarithmic growth rate of matter fluctuations, and 
$\mu_k$ is the cosine of the angle between ${\bf k}$ and the LOS. 

\subsection{Legendre Moments of $\xi({\bf r})$}

In linear theory (Eqn.~\ref{eq:kaiser}), only the $\ell=0,2$ and $4$ 
moments contribute to the power spectrum $P_g^s({\bf k})$, and its Fourier transform
$\xi_g^s({\bf s})$.
The two are simply related by $\ell^{th}$ order Bessel functions:
\begin{equation}
\xi_{\ell}(s) = i^{\ell} \int \frac{k^2 dk}{2\pi^2} P_{\ell}(k) j_{\ell}(ks).
\end{equation}

Given a tight constraint on the underlying {\em shape} of the linear 
matter power spectrum, the
two-dimensional clustering of galaxies constrains both $b\sigma_8$ and
$f\sigma_8$ \citep{PerWhi08,WhiSonPer09}.
In this work we measure and model only the monopole and quadrupole moments
of the correlation function, $\xi_{0,2}(s)$.
These two moments are sufficient to constrain both $b\sigma_8$ and
$f\sigma_8$ and encompass most of the available information on
the peculiar velocity field for the highly biased galaxies of interest
here \citep{ReiWhi11}, in addition to being more easily modelled than
higher-$\ell$ moments \citep[but see also][]{TarSaiNis11}.  
Therefore, we collapse the anisotropic clustering
information in $\xi(r_{\sigma},r_{\pi})$ into two one-dimensional functions
$\xi_{0,2}(s)$ before extracting cosmological information from the
anisotropic galaxy clustering.

\subsection{Redshift space halo clustering in the quasilinear regime}

The Kaiser formula describing the linear effect of redshift space
distortions breaks down even on quite large scales.  An accurate model
of the two-dimensional clustering of galaxies must account for nonlinear
evolution in the real space matter density and velocity fields,
nonlinear galaxy bias, and the nonlinear mapping between real and
redshift space separations of pairs of galaxies.
The simplest picture of galaxy formation asserts that galaxies occupy
dark matter halos, and so as a step towards understanding the clustering
of galaxies, \citet{ReiWhi11} showed that a
streaming model where the pairwise velocity probability distribution function is
approximated as Gaussian can be used to relate real space clustering and pairwise
velocity statistics of halos to their clustering in redshift space.  
We will demonstrate 
in Sec.~\ref{halos2gals} that the same model describes the clustering of galaxies:
\begin{equation}
  1+\xi^{s}_{\rm g}(r_{\sigma},r_{\pi}) = \int \left[1+\xi^{r}_{\rm g}(r)\right]
  e^{-[r_\pi - y - \mu v_{12}(r)]^2/2\sigma_{12}^2(r,\mu)} \frac{dy}{\sqrt{2\pi\sigma^2_{12}(r,\mu)}} \label{streamingeqn},
\end{equation}
\noindent
where $r_\sigma$ and $r_\pi$ are the redshift space transverse and LOS distances
between two objects with respect to the observer, $y$ is the {\em real} space
LOS pair separation, $\mu = y/r$, $\xi_{\rm g}^{\rm r}$ is the
real space galaxy correlation function, $v_{12}(r)$ is the average infall velocity of
galaxies separated by real-space distance $r$, and $\sigma_{12}^2(r,\mu)$ is the
rms dispersion of the pairwise velocity between two galaxies separated with
transverse (LOS) real space separation $r_{\sigma}$ ($y$).  

$\xi_{\rm g}^{\rm r}(r)$, $v_{12}(r)$ and $\sigma_{12}^2(r,\mu)$ are computed
in the framework of Lagrangian ($\xi^{\rm r}$) and standard perturbation
theories ($v_{12}$, $\sigma_{12}^2$).  
Only two nuisance parameters are necessary to describe the clustering of a
sample of halos or galaxies in this model: $b_{1L} = b-1$, the first-order
Lagrangian host halo bias in {\em real} space, and $\sigma^2_{\rm FoG}$, an
additive, isotropic velocity dispersion accounting for small-scale motions
of halos and galaxies which will be described below.
Further details of the model, its numerical implementation, and its accuracy
can be found in \citet{ReiWhi11} and Appendix \ref{cosmoxi2d}. 

\subsection{From halos to galaxies}
\label{halos2gals}

\citet{ReiWhi11} examined the validity of Eqn.~\ref{streamingeqn} only for halo
clustering rather than galaxies, thus the model must be extended and checked
with a realistic sample of mock galaxies.  
We use the machinery of the halo model \citep[see][for a review]{CooShe02}
to describe the galaxy density field in terms of the density field of the
host halos.
Of particular importance for modeling redshift space distortions is the
distinction between ``central'' and ``satellite'' galaxies \citep{Kau93,Col94,Kra04}.
When modeling an approximately mass-limited galaxy sample, the first galaxy
assigned to a host halo is considered central, and its position and velocity
are that of the host halo center.
Satellite galaxies orbit in the potential well of the host halos, and so are
offset in both position and velocity from the halo center.  In our particular
implementation of the halo model, satellite galaxies are randomly drawn from the dark matter
particle members of the host halo in our simulation.  As the virial
velocities of massive halos can be large (amounting to redshift space LOS
separations of tens of $h^{-1}$ Mpc), intrahalo velocities can distort the redshift 
space correlation function.  In the limit that these virial motions are
uncorrelated with the quasilinear velocity field of interest, they can be
accounted for by additional convolution along the line of sight.  

\begin{figure}
\begin{center}
\resizebox{3.2in}{!}{\includegraphics{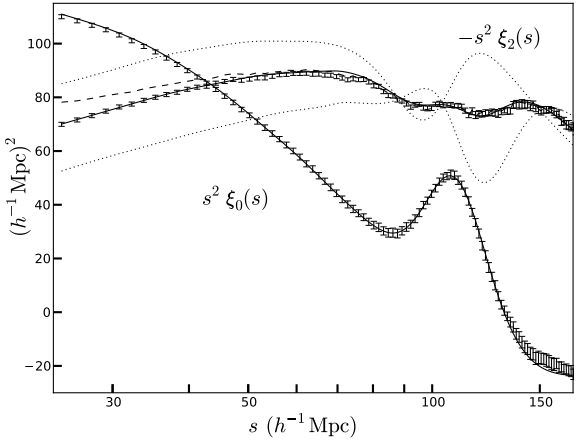}}
\end{center}
\caption{Error bars enclose the mean $s^2\xi_{0,2}(s)$ measured from the
\citet{Whi11} mock galaxy catalogs.  The solid line is our model fit, where
$\sigma^2_{\rm FoG}$ has been varied to minimize the difference.  The dashed line
shows $s^2 \xi_2$ for the mock galaxies when their intrahalo velocities are
artificially set to 0, and indicates that intrahalo velocities suppress $\xi_2$
by $\approx 10$ per cent on the smallest scales we are fitting.  The nuisance parameter
$\sigma^2_{\rm FoG}$ adequately describes the effect of intrahalo velocities.
Dotted lines show the predicted $\xi_2$ when varying the Alcock-Paczynski parameter
$F(z)$ by $\pm 10$ per cent and holding $D_V$ (and thus $\xi_0$) fixed.}
\label{fig:theoryfits}
\end{figure}

To assess the impact of virial motions on the observed galaxy clustering we
use the mock catalogs described in \cite{Whi11}, which closely match the
small-scale clustering of CMASS galaxies.
We compute $\xi_{0,2}(s)$ from these mocks
(the average is shown as the error bars in Figure \ref{fig:theoryfits})
and recompute $\xi_{0,2}$ after artificially setting the intrahalo velocities
to 0 (the dashed curves, with the one for $\xi_0$ covered by the solid line).
Intrahalo velocities suppress the amplitude of $\xi_2$ on the smallest scales
we attempt to model, reaching a 10 per cent correction at $25\,h^{-1}$Mpc.
The reason for this suppression is that, on small scales, $d\xi^r/dy<0$
and non-negligible.  This causes a net transport of pairs to larger separations
in redshift space.
Note this is opposite to the effect of quasilinear peculiar velocities, which
make the separation of a pair in redshift space on average smaller than in real
space.

We include the effect of intrahalo velocities in our model by including an
extra convolution with a Gaussian of dispersion $\sigma_{\rm FoG}$.
The solid curves in Fig.~\ref{fig:theoryfits} are then the predictions
for $\xi_{0,2}$ with the best fit $\sigma^2_{\rm FoG}=(3.2\,h^{-1}{\rm Mpc})^2$
or $21\,{\rm Mpc}^2$.  The model successfully describes the effect of intrahalo
velocities on the monopole and quadrupole correlation functions.

In Appendix \ref{galaxydetails} we quantify the impact of our uncertainties in
the halo occupation distribution (HOD) of CMASS galaxies and the
possible breakdown of our assumption that the first galaxy assigned to each
halo is ``central'' (i.e., has no intrahalo velocity dispersion) on the value of
our nuisance parameter $\sigma^2_{\rm FoG}$.  As a result of these considerations,
we place a uniform prior on $\sigma^2_{\rm FoG}$ between 0 and 40 (Mpc)$^2$. 

\subsection {Alcock-Paczynski Effect}
\label{APsec}

Galaxy redshift surveys collect two angular coordinates and a redshift for each
galaxy in the sample.  A fiducial cosmological model must be adopted to generate
maps and measure clustering as a function of comoving separations.
This mapping depends on both the angular diameter distance and the inverse of
the Hubble parameter at the redshift of each galaxy pair.
To a good approximation the inferred galaxy clustering in a different
cosmological model can be obtained from the fiducial one by a single rescaling
of the transverse and parallel separations \citep{Per10}.
Rather than modify our observed galaxy clustering, we will account for the AP
effect when we test different cosmological models by introducing two scale
parameters, $\alpha_{\perp}$ and $\alpha_{\parallel}$, into the theoretical
correlation functions we are fitting to:
\begin{eqnarray}
\xi^{\rm fid}(r_\sigma, r_\pi) & = & \xi^{\rm true}(\alpha_{\perp} r_{\sigma},
\alpha_{\parallel} r_{\pi}), \label{eq:APeqn}\\
\alpha_{\perp} = \frac{D_{A}^{\rm fid}(z_{\rm eff})}{D_{A}^{\rm true}(z_{\rm eff})},
&\;& \alpha_{\parallel} = \frac{H^{\rm true}(z_{\rm eff})}{H^{\rm fid}(z_{\rm
eff})},
\end{eqnarray}
where $D_A$ denotes the physical angular diameter distance. 
Here $\xi^{\rm true}$ is the expected two-dimensional correlation function 
{\em if} the measured galaxy correlation function were computed 
 assuming the true redshift-distance relation in the cosmology being tested. 
$\xi^{\rm fid}$ is the prediction for the measured correlation function, given that  
galaxy separations were computed using the fiducial cosmology model.  
That is, $\alpha_{\perp}$ and $\alpha_{\parallel}$
scale the `true' separations to the ones calculated using the fiducial
cosmology.

The spherically averaged correlation function, $\xi_0$, is sensitive to the parameter
combination 
\begin{equation}
D_V(z) \equiv \left((1+z)^2 D_A^2(z)
\frac{cz}{H(z)}\right)^{1/3}.
\label{eq:DV}
\end{equation}
The quadrupole of the measured correlation
function allows a measurement of a second combination \citep{AP,PadWhi08,Bla11c}
\begin{equation}
F(z) = (1+z) D_A(z) H(z)/c,
\label{eq:FAP}
\end{equation}
thus breaking the degeneracy between $(1+z_{\rm eff}) D_A$ and $H$.  To a
good approximation, changing $D_V$ simply rescales the value of $s$ in the
predicted correlation function, while $F(z)$ primarily affects the
quadrupole.  Figure \ref{fig:theoryfits} shows the effect of varying $F$ 
by $\pm 10$ per cent on $\xi_2$ at fixed $D_V$ with the dotted curves.  The
scale-dependence of $\Delta \xi_2$ due to the AP effect will allow us to
separate the effects of peculiar velocities and the AP effect.
\section {Analysis}
\label{sec:analysis}

\subsection{Cosmological Model Space}
\label{cosmomodel}
Given an underlying linear matter power spectrum shape $P_{\rm lin}(k, z_{\rm eff})$, we
consider the predicted galaxy clustering as a function of five parameters:
$\vec{p}_{\rm gal} = \{b\sigma_8, f\sigma_8, \sigma_{\rm FoG},  (1+z_{\rm eff}) D_A(z_{\rm eff}),
H(z_{\rm eff})\}$.  Since the normalization of $P_{\rm lin}(k,z_{\rm eff})$ (denoted
throughout as $\sigma_8$) determines the amplitude of the second-order
perturbation theory corrections in our model, in principle we should be able to
separately determine $b\sigma_8$, $f\sigma_8$, and $\sigma_8$.  In practice, the
dependence is sufficiently small and degenerate with the nuisance parameters
$b\sigma_8$ and $\sigma^2_{\rm FoG}$ that the degeneracy cannot be broken; see
Appendix \ref{sec:modelacc} for details. 

We have also assumed that any error in the fiducial cosmological model
used to compute $\xi_{0,2}(s)$ from the CMASS galaxy catalog can be absorbed in a
single scaling of distances, interpreted at the effective redshift of the
survey.  If the assumed redshift dependence of $(1+z_{\rm eff}) D_A(z)$ and $H(z)$ is grossly
incorrect, we would expect a difference in the correlation functions split on
redshift; we see no evidence for this in our tests \citep{Ross:2012}.

\subsubsection{Prior on the linear matter power spectrum from WMAP7}

While the scale dependence of galaxy clustering itself can constrain the shape
of the linear matter power spectrum, at present (and certainly with the imminent
public release of Planck data) the constraints enabled by CMB measurements are stronger.
The strong CMB constraints mean we can use the entire linear matter power
spectrum as a standard ruler determined by observations of the CMB, rather
than only the BAO feature.  This approach relies on further cosmological model
assumptions that are consistent with the current data, but from which moderate
deviations are still allowed.

The temperature of the CMB has been measured exquisitely well \citep{Mather94},
and determines the physical energy density in radiation, $\Omega_r h^2$.  In the
minimal cosmological model allowed by current observations \citep{WMAP7}, namely
a flat $\Lambda$CDM cosmology with nearly scale-invariant scalar, adiabatic,
Gaussian fluctuations along with the three standard, nearly massless neutrino
species,  
only three additional parameters determine the shape of
the underlying linear matter power spectrum,  $P_{\rm lin}(k)$.   Relative peak
heights in the CMB determine the physical energy densities in cold and baryonic
matter, $\Omega_{c,b} h^2$, and the overall scale-dependence of the CMB power
spectrum determines the spectral index $n_s$ of the nearly scale-invariant
scalar primordial fluctuations.  Constraints on these parameters do
not rely on the distance to the cosmic microwave background, and thus are immune
to the behavior of dark energy at lower redshifts than the last scattering
surface.  Moreover, small-scale CMB experiments \citep{Kei11,Hlo11} now probe
fluctuations on the same scales as galaxy clustering measurements, and find no
compelling evidence for, e.g., a running of the spectral index.  
Allowing for running of the spectral index would degrade the CMB constraints
on the linear matter power spectrum \citep[e.g.,][]{Mehta12} but we will not
include this additional parameter--obviously, our constraints should be
interpreted in the context of our model assumptions.

One important extension of the minimal cosmological model is allowing neutrinos to have mass;
neutrino oscillation experiments suggest that $\sum m_{\nu} \gtrsim 0.05\,$eV
\citep{abazajian:2011}.  As the universe expands and cools, massive neutrinos become
non-relativistic and modify the linear matter power spectrum inferred from the
CMB, as well as alter the expansion history as compared with the massless
neutrino case.  In this work we ignore this additional uncertainty in the shape
of $P_{\rm lin}(k)$. Current upper bounds that combine several cosmological probes
find $\sum m_{\nu} \lesssim 0.3$ eV \citep[e.g.,][]{Rei10b,dePutter:2012}, which
is safely below the detectable level in the DR9 CMASS sample \citep{Sanchez:2012}.

As we explore the cosmological constraints available from our data set, we will
consider a set of power-spectrum shapes parametrized by $\vec{p}_{\rm
s}=\{\Omega_{\rm b}h^2, \Omega_{\rm c}h^2, n_{\rm s}\}$. We will marginalize
over parameters $\vec{p}_{\rm s}$ either by importance resampling
\citep{MCMC,lewisMCMC} the public Monte Carlo Markov Chains for a spatially flat
$\Lambda$CDM model provided by the WMAP collaboration, or by approximating the
constraints on $\vec{p}_{\rm s}$ with a three-dimensional Gaussian likelihood.
Note that the three-dimensional constraints on linear matter power spectrum
parameters are very nearly independent of cosmological model extensions that
change the expansion rate only for $z \ll 1000$ (i.e., introducing $w$ or $\Omega_k$).

\subsection{Models}
\label{models}
In the present paper, we interpret the anisotropic clustering of CMASS galaxies
only in the context of the $\Lambda$CDM cosmology.  By relaxing assumptions
about the redshift-distance relation and/or the growth of structure in the
$\Lambda$CDM model, we report the statistical precision with which CMASS
measurements constrain both the peculiar velocity field $f\sigma_8(z_{\rm eff})$
and geometric quantities $D_{A}(z_{\rm eff})$ and $H(z_{\rm eff})$, without adopting a
particular cosmological model that specifies how these quantities are related.
We consider the following four models:
\begin{itemize}
  \item {\bf Model 1: WMAP7+CMASS flat $\Lambda$CDM.}  In the flat $\Lambda$CDM
cosmology, $f\sigma_8(z_{\rm eff})$, $D_{A}(z_{\rm eff})$ and $H(z_{\rm eff})$ are all
determined once $\Omega_m$, $\sigma_8$, and $H_0$ are specified.  This provides
a null test of our assumptions relating CMB fluctuations to predicted galaxy
fluctuations.
  \item {\bf Model 2 WMAP7+CMASS $\Lambda$CDM geometry, free growth.}  The
analysis for this model is the same as Model 1, except that we consider
$f\sigma_8$ a free parameter in the CMASS galaxy clustering fits.
  \item {\bf Model 3 WMAP7+CMASS $\Lambda$CDM growth, free geometry}  The
analysis for this model is the same as Model 1, except that we consider
$D_{A}(z_{\rm eff})$ and $H(z_{\rm eff})$ as free parameters in the CMASS galaxy
clustering fits.    
  \item {\bf Model 4 WMAP7+CMASS, free growth, free geometry}: In this model,
$f\sigma_8(z_{\rm eff})$, $D_{A}(z_{\rm eff})$ and $H(z_{\rm eff})$ are all free parameters
in the fit to the galaxy clustering data.  This allows us to determine how well
our data break the degeneracy between the RSD and AP effects, and to present the
constraints originating from the amplitude and scale-dependence of the galaxy 
quadrupole $\xi_2(s)$ in the most model independent way possible.  This
multivariate distribution can be used to constrain any model that does not alter
the shape of the linear matter power spectrum at $z_{\rm eff}$ from that
inferred from the CMB on the relatively large scales of interest here.
\end{itemize}   
In all four scenarios we allow the value of $\sigma^2_{\rm FoG}$ to vary between
$0\ {\rm Mpc}^2$ and $40\ {\rm Mpc}^2$ with a flat prior and marginalize over
both $\sigma^2_{\rm FoG}$ and $b\sigma_8$ when deriving final results.  
We detail our methods for sampling the multi-dimensional probability
distribution functions of interest in
Models 1 through 4 in Appendix \ref{sec:mcmcmethods}. 

This approach to parameter fitting allows our estimates of the growth of
structure and geometry to be independent of many model assumptions and can be
used to put constraints on more general models of gravity and dark energy. However,
they still rely on the standard model in three ways. First, we assume that
the processes in the early Universe that were responsible for setting up the
linear matter power spectrum at recombination do not change significantly, which is
true for the most popular models of modified gravity and dark energy. Second, we
assume that growth at the level of linear perturbation theory is scale
independent between the CMB epoch and the effective redshift of our sample.  
Third, we use GR to
compute the perturbation theory corrections to the galaxy clustering
predictions. The perturbation theory corrections are not large, and they are most
important on small scales where $\sigma^2_{\rm FoG}$ also becomes important;
therefore we cannot strongly constrain the amplitude of the higher-order
corrections.  Constraints on models with scale-dependent growth should be
derived directly from the correlation function measurements and their covariance.

\subsection{The meaning of $\sigma_8$}
\label{sigma8def}
We follow the standard convention of denoting the amplitude of the matter
power spectrum by $\sigma_8^2$, even though we restrict our analysis to scales $s >
25$ $h^{-1}$ Mpc, so a different weighted integral over $P(k)$ that is
concentrated on larger scales would more accurately reflect 
our constraints on the growth rate of matter fluctuations.
In particular, since $P(k)$ is well-determined for $k$ in Mpc$^{-1}$
\citep[e.g.][]{Whi06}, and the BAO scale provides a standard ruler with
percent-level precision, our data constrain the amplitude of matter
fluctuations on scales $\gtrsim 36\,$Mpc.
In practice, the tight constraints on the shape of $P(k)$ means that 
differences arising from how one specifies its amplitude  
are small when computing parameter
constraints, as long as $h$ is well-determined in the model.  In Model 2, we
sample power spectra from the WMAP $\Lambda$CDM chain, and take the traditional
value of $\sigma_8$ to relate the model parameter $f$ and the reported
constraint $f\sigma_8$.  In Model 4, we do not specify a value of $h$ with each
sampled power spectrum, so we normalize the power spectra by fixing
$\sigma_{R}$, where $R=8/0.7= 11.4\,$Mpc.  For power spectra drawn from WMAP7
$\Lambda$CDM chains, $\sigma_R/\sigma_8=0.99\pm 0.024$; the offset and
variance between these parameters is negligible compared to our measurement
errors on $f\sigma_8$.

\section{Results}
\label{sec:results}
\begin{table*}
\begin{center}
\begin{tabular}{lllllll}
Model & $b\sigma_8$ & $f\sigma_8$ & $D_V$ [Mpc] & $F$ &
$(1+z_{\rm eff}) D_A$ [Mpc] & $H$ [km s$^{-1}$ Mpc$^{-1}$] \\
\hline
2 & $1.228_{-0.032}^{+0.033}$ & $ 0.415_{-0.033}^{+0.034} $ & - & - & - & - \\
3 & $1.246^{+0.043}_{-0.046}$ & - & $2076^{+42}_{-44}$ & $0.683^{+0.026}_{-0.025}$ & $2204 \pm 44$ & $92.9^{+3.6}_{-3.3}$ \\ 
4 & $1.238^{+0.047}_{-0.050}$ & $0.427^{+0.069}_{-0.063}$ & $2070^{+43}_{-46}$ & $0.675^{+0.042}_{-0.038}$ & $2190 \pm 61$ & $92.4^{+4.5}_{-4.0}$\\
WMAP7 $\Lambda$CDM & - & $0.451\pm0.025$ & $2009 \pm 42$ & $0.6635^{+0.0084}_{-0.0073}$ & $2113^{+53}_{-52}$ & $94.2^{+1.4}_{-1.3}$ \\
\end{tabular}
\end{center}
\caption{
The median and 68.3 per cent confidence level intervals on parameters $b\sigma_8$,
$f\sigma_8$, absolute distance scale $D_V$ (Eqn.~\ref{eq:DV}), Alcock-Paczynski
parameter $F$ (Eqn.~\ref{eq:FAP}), as well as derived parameters, comoving
angular diameter distance ($(1+z_{\rm eff}) D_A$) and expansion rate ($H$).
To obtain these constraints, we marginalize over $\sigma^2_{\rm FoG}$ and power
spectrum shape parameters $\vec{p}_{\rm s}=\{\Omega_{\rm b}h^2, \Omega_{\rm
c}h^2, n_{\rm s}\}$ for Models 2-4, as described in Section \ref{models}.  We
interpret our measurements at the effective redshift of our galaxy sample,
$z_{\rm eff} = 0.57$.
}
\label{tab:bf4b4}
\end{table*}
In this section we present the results of fitting our analytic model for
$\xi_{0,2}(s)$ to the observed galaxy correlation functions.  
Figure \ref{fig:DVvsomh2} summarizes our constraints from the shape of the observed
angle-averaged correlation function $\xi_0(s)$, while Figures \ref{fig:onedhist}
and \ref{fig:twodhist} highlight our parameter constraints from the observed 
anisotropic galaxy clustering.
Constraints on both the peculiar velocity amplitude and geometric quantities
$(1+z_{\rm eff}) D_A(z_{\rm eff})$ and $H(z_{\rm eff})$ are summarized in Table \ref{tab:bf4b4} for Models
2-4.  

\subsection{Goodness of fit and $\Lambda$CDM results}
\label{sec:goodfitLCDM}

We include 23 separation bins for both $\ell=0$ and $\ell=2$ in our $\chi^2$
analyses.  In this section, we consider models with increasing numbers of
free parameters, and ask whether changes in $\chi^2$ across the models indicate a
preference for parameter values outside the predicted values from WMAP7 in the
$\Lambda$CDM model.  We first fix the underlying power spectrum to the one
assumed for all of the mock catalogs, and also fix $\sigma^2_{\rm FoG}=21\,{\rm Mpc}^2$,
the best fit value to our $N$-body based mock galaxy catalogs.
We vary the galaxy bias, and find a minimum $\chi^2=45.7$ for 45 degrees of
freedom, demonstrating that the mock galaxy catalogs used to validate our model
and compute our covariance matrix are consistent with the observed galaxy
clustering.
If we allow $\sigma^2_{\rm FoG}$ to vary as well, $\chi^2=42.1$ at
$\sigma^2_{\rm FoG}=40\,{\rm Mpc}^2$.
The difference indicates that we cannot expect a strong constraint on
$\sigma^2_{\rm FoG}$ within our prior when other cosmological parameters are
varying; it is important, however, to marginalize over this nuisance parameter,
since it increases our uncertainty in $f\sigma_8$; see the discussion in
Section \ref{errorbudget}.

If we restrict ourselves to $\Lambda$CDM models consistent with WMAP7
(Model 1), we find a minimum $\chi^2$ value of 39.3 at
$\Omega_m h^2 = 0.1395$ and $H_0 = 68.0\,{\rm km}\,{\rm s}^{-1}{\rm Mpc}^{-1}$
when $b\sigma_8$ and $\sigma^2_{\rm FoG}$ are varied.
Model 4 has the most free parameters: five describing the galaxy clustering and
three specifying the linear matter power spectrum.  In this case we find a
minimum $\chi^2$ of 39.0 for 41 degrees of freedom.  This best-fitting model is
shown with our measurements of the correlation function in 
Figures \ref{fig:butterfly} and \ref{fig:xilmeas}, and has
parameter values $b\sigma_8 = 1.235$, $f\sigma_8 = 0.437$, $\sigma_{\rm FoG} = 40$
Mpc$^{2}$, $D_A = 2184$ Mpc, $H = 91.5\,{\rm km}\,{\rm s}^{-1}{\rm Mpc}^{-1}$, 
$\Omega_m h^2 = 0.1364$, $\Omega_b h^2 = 0.02271$, $n_s = 0.967$. 
We conclude that the observed $\xi_{0,2}$ is fully consistent with the
$\Lambda$CDM cosmology; changes in $\chi^2$ values between the models do not
indicate a significant preference for parameter values of $f\sigma_8$,
$F$, and $D_V/r_s$ outside of the values predicted by WMAP7 in the
$\Lambda$CDM model.

CMASS measurements also improve constraints compared to
WMAP7 in the $\Lambda$CDM model: 
$\Omega_m h^2 = 0.1363 \pm 0.0035$, $\Omega_m = 0.283 \pm
0.017$, and $H_0=(69.3\pm1.5){\rm km}\,{\rm s}^{-1}{\rm Mpc}^{-1}$;
WMAP7 alone finds
$\Omega_m h^2 = 0.1334 \pm 0.0056$, $\Omega_m = 0.266 \pm 0.029$, and
$H_0 =(71.0 \pm 2.5){\rm km}\,{\rm s}^{-1}{\rm Mpc}^{-1}$.  Comparison with the BAO-only results of
\citet{Aardvark} demonstrates that in this minimal model, nearly all of the
additional information
on these three parameters is coming from the BAO feature.

\begin{figure}
  \includegraphics[width=85mm]{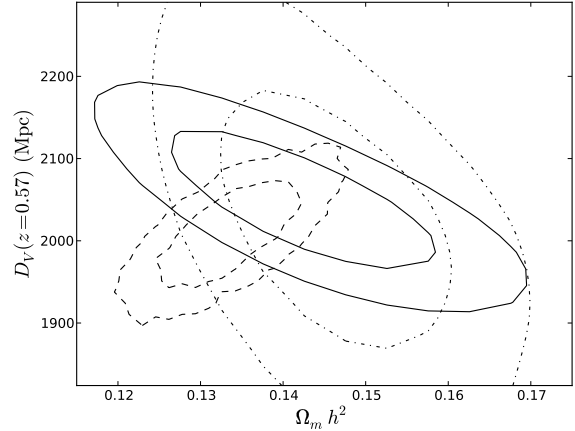}
  \caption{Contours of $\Delta \chi^2 = 2.30$ and $6.17$ for fixed $\Omega_b h^2 = 
0.02258$ and $n_s = 0.963$ for the monopole ($\ell = 0$) galaxy clustering measurements alone (solid).
For comparison, we also compute $\chi^2$ using the ``no-wiggles'' power spectrum
from \citet{EisHu98} (dash-dot) to isolate information from the broadband shape of
the correlation function without the BAO feature; this fit is primarily sensitive to the
apparent location of the peak in $P(k)$, which corresponds to the horizon size at 
matter-radiation equality, $\propto \Omega_m h^2
D_V(z_{\rm eff})$.  We also project the WMAP $\Lambda$CDM constraints onto these
parameters, and show 68 and 95 per cent contours (dashed).}  
  \label{fig:DVvsomh2}
\end{figure}

However, the shape of the measured galaxy correlation function does provide an
independent probe of the underlying linear matter power spectrum.  
With a strong prior on $\Omega_b h^2$ and $n_s$ taken from the CMB, the clustering of galaxies is
sensitive to the peak in the linear matter power spectrum, which depends on the
horizon size at matter-radiation equality, $\propto \Omega_m h^2$ at fixed
effective number of relativistic species ($N_{\rm eff} = 3.04$ for the
standard three neutrino species).  
The scale at which the peak appears depends on the low
redshift distance relation, so the broadband shape of the angle-averaged 
galaxy power spectrum or correlation function constrains $\Omega_m h^2
D_V(z_{\rm eff})$, in addition to the constraint on $D_V(z_{\rm eff})/r_s(z_{\rm drag})$
that comes from the location of the BAO feature.
We illustrate the constraining power of our dataset by fixing $\Omega_b h^2 =
0.02258$ and $n_s = 0.963$, and computing the CMASS-only likelihood in the
$D_V-\Omega_m h^2$ plane shown in Figure \ref{fig:DVvsomh2} as the solid
contours.  
For this exercise we use only the monopole ($\ell = 0$) measurements, and find a
minimum $\chi^2$ value of 18.2 for 19 degrees of freedom (DOF).

For comparison, we also isolate the broadband shape information by fitting to a
no-wiggle power spectrum \citep{EisHu98}, which should primarily
be sensitive to $\Omega_m h^2 D_V(z_{\rm eff})$.  Results of this fit are shown
as dash-dot contours in Figure \ref{fig:DVvsomh2}.  This model provides a poor fit
to the measured correlation function, with $\chi^2_{min} = 38.5$ for 19 DOF, indicating a
strong preference for models with the expected BAO feature. However, we do find
that the inferred value of $\Omega_m h^2$ from the broadband shape of the 
measured galaxy correlation function is consistent with the prediction from the CMB.

Projecting the with-BAO model fits onto
$\Omega_m h^2$, we find $\Omega_m h^2 = 0.142 \pm 0.011$.  Translating the
results of a similar analysis from \citet{Rei10} for the SDSS-II LRG sample 
($z_{\rm eff} = 0.31$) to the same assumptions yields $\Omega_m h^2 =
0.141^{+0.010}_{-0.012}$, while the WiggleZ analysis of emission line galaxies
at $z_{\rm eff} \approx 0.6$ finds $\Omega_m h^2 = 0.127 \pm 0.011$ \citep{Bla11a}.
Strictly speaking, these constraints are not uncorrelated since they have a
small amount of overlapping volume;
neglecting their small correlation, the combined galaxy clustering estimate for
$\Omega_m h^2$ is $0.137 \pm 0.0064$, which marginalizes over the low redshift
distance-redshift relation and is in excellent agreement with the 
WMAP7 $\Lambda$CDM constraint of $\Omega_m h^2 = 0.1334 \pm 0.0056$
(dashed contours in Figure \ref{fig:DVvsomh2}).

\subsection{Constraints on the peculiar velocity field amplitude}
\begin{figure*}
  \includegraphics[width=170mm]{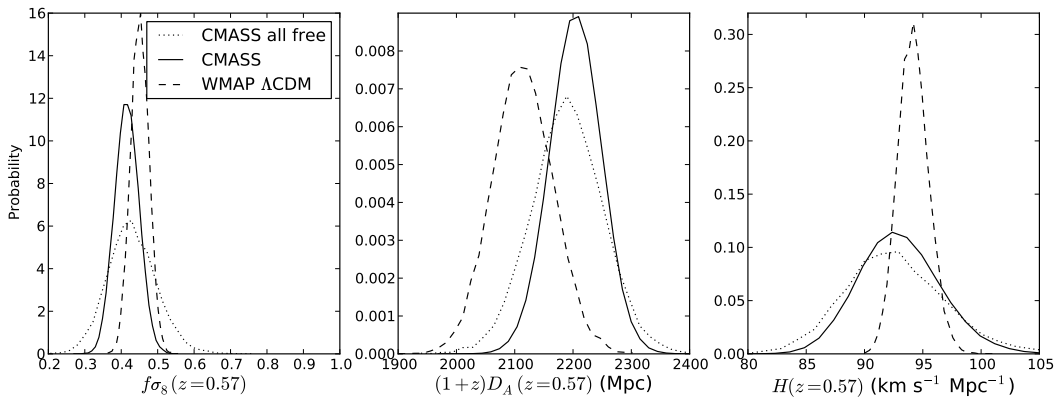}
  \caption{One-dimensional constraints on $f\sigma_8$, $(1+z_{\rm eff}) D_A(z_{\rm eff})$, and
$H(z_{\rm eff})$ under different model assumptions.  
The dashed curves indicate WMAP7-only $\Lambda$CDM.  The solid
(models 2 and 3) and dotted (model 4) curves are constraints derived from the
CMASS $\xi_{0,2}$ measurements with a WMAP7 prior on the underlying linear
matter power spectrum $P(k/{\rm Mpc}^{-1})$.  The solid curves additionally use
the $\Lambda$CDM parameters in the WMAP7 chains to fix either $(1+z_{\rm eff}) D_A(z_{\rm eff})$
and $H(z_{\rm eff})$ (left panel), or $f\sigma_8$ (right two panels).  All three
constraints degrade when fitting for geometry and growth simultaneously using
the CMASS observations.}  
  \label{fig:onedhist}
\end{figure*}
In Figures \ref{fig:onedhist} and \ref{fig:twodhist} we compare our constraints on the peculiar
velocity field amplitude, $f\sigma_8(z_{\rm eff})$, from Models 2 (solid) and 4
(dotted) to the predicted distribution from WMAP7 (dashed), assuming a
flat $\Lambda$CDM cosmology.  The 68 per cent confidence intervals for Models 2 and 4 are
listed in Table \ref{tab:bf4b4}.  These measurements agree with the
$\Lambda$CDM WMAP7 expectation, $0.451 \pm 0.025$.

\subsection{Geometric constraints}
Our tightest geometric constraint comes from the BAO feature in the monopole
correlation function.  The cosmological parameter dependence of the location of
the BAO feature is given by the sound horizon at the drag epoch, $r_s(z_{\rm
drag})$ \citep[we use the definition in][]{EisHu98}.
We find
($D_V(z_{\rm eff})/r_s(z_{\rm drag})/(D_V(z_{\rm eff})/r_s(z_{\rm drag}))_{\rm fiducial} = 1.023 \pm 0.019$.
The difference between our best fit value and the pre-reconstruction fits to
monopole correlation function presented in \citet{Aardvark} and 
 \citet{Sanchez:2012} 
is due to our different choice of binning rather than our fitting
methodology; we verified that with the
same measurement values and covariance matrix, our method recovers the same central
value as the result presented in \citet{Aardvark}. 

Though our fits include information from the broadband shape of the
correlation function, the resulting central value and error on $D_V/r_s(z_{\rm drag})$ are
consistent with the fits performed in \citet{Aardvark}, which marginalize
over the broadband shape of correlation function or power spectrum.  We therefore
conclude that essentially {\em all} of the information on the distance scale
$D_V$ is coming from the BAO feature in the correlation function, as was also
true in the analysis of the SDSS-II LRG power spectrum \citep{Rei10}.  This can
be seen in Fig.~\ref{fig:DVvsomh2}, where at fixed $\Omega_m h^2$, the
constraint on $D_V$ is 2.5 times weaker for the ``no-wiggles'' fit compared to
the fit including the BAO feature.  In addition, our central value and error on 
$D_V/r_s$ are consistent when we fit $\xi_0$ only, or $\xi_0$ and $\xi_2$ simultaneously.  

Finally, we note that many of the small differences between the cosmological constraints
presented here and those in our companion papers stem from slight
differences in the best fitting value for $D_V/r_s(z_{\rm drag})/(D_V(z_{\rm
eff})/r_s(z_{\rm drag}))_{\rm fiducial} $.  The
correlation function and power spectrum post-reconstruction ``consensus'' value from
\citet{Aardvark} is  $1.033 \pm 0.017$; this value was used in cosmological parameter
studies in that paper.  \citet{Sanchez:2012} found $1.015 \pm 0.019$, in
agreement with the pre-reconstruction analysis of the correlation function
presented in \citet{Aardvark}.

Once the WMAP7 prior on the underlying linear matter power spectrum $p_s$ is
included and information from the Alcock-Paczynski effect is included through
$\xi_2$, the standard ruler from the CMB allows us to infer $(1+z_{\rm eff}) D_A(z_{\rm eff})$ and
$H(z_{\rm eff})$ separately.  Constraints from Model 3 (solid) and Model 4
(dotted) on $(1+z_{\rm eff}) D_A(z_{\rm eff})$ and $H(z_{\rm eff})$ are shown in
Figures \ref{fig:onedhist} and \ref{fig:twodhist}.  
Model 3 further uses the WMAP7 $\Lambda$CDM prediction for
$f\sigma_8$ to disentangle the RSD and AP effects, and we find factors of 1.3
and 1.2 improvement in $(1+z_{\rm eff}) D_A$ and $H$ errors when adopting this additional
assumption: $(1+z_{\rm eff}) D_A = 2204 \pm 44$ ($2190 \pm 61$) Mpc, $H =
92.9^{+3.6}_{-3.3}$ ($92.4^{+4.5}_{-4.0}$) km s$^{-1}$ Mpc$^{-1}$.  In both
models, the CMASS distance constraints are consistent with what is inferred from
WMAP7 alone in a $\Lambda$CDM cosmology: $(1+z_{\rm eff}) D_A(z_{\rm eff}) =
2113^{+53}_{-52}$, $H(z_{\rm eff}) = 94.2^{+1.4}_{-1.3}$.  We compare the
two-dimensional constraints on $(1+z)D_A$ and $H$ from Models 3 and 4 with the
prediction from WMAP7 for a flat $\Lambda$CDM model in Fig.~\ref{fig:twodhist},
which shows that CMASS constraints on $(1+z)D_A$ and $H$ are only weakly correlated.
\subsection{Using our results}
Our results may be used to test cosmological models which share the assumptions
we have adopted in this analysis.  Most importantly, we have assumed adiabatic
and scale-invariant primordial fluctuations, and that the transfer function was computed
assuming the standard number of massless neutrino species, $N_{\rm eff} = 3.04$.
We have assumed that the linear growth is scale-independent,
and account for non-linear corrections using perturbation theory within general
relativity.
The code to evaluate our theoretical prediction as a function of the underlying
linear matter power spectrum, {\bf cosmoxi2d} is publicly
available\footnote{http://mwhite.berkeley.edu/CosmoXi2D}.  
For most purposes, however, our results can be well-approximated by the following 
multivariate Gaussian likelihood for the parameters 
$p_{3d} = \{f\sigma_8, F, (D_V/r_s)/(D_V/r_s)_{\rm
fiducial}\}$, which should be interpreted at $z_{\rm eff} = 0.57$:
\begin{equation} \label{gaussapprox}
\bar{p}_{3d} = 
\begin{pmatrix}
0.4298 \\
0.6771 \\
1.0227
\end{pmatrix}
\end{equation}
\begin{equation}
10^3 {\bf C} = 
\begin{pmatrix}
  4.509868 & 2.435891 & -0.01087251 \\
2.435891 & 1.736087  & -0.06287155 \\
-0.01087251 & -0.06287155 & 0.3548373
\end{pmatrix}
\end{equation}
Deviations from the Gaussian likelihood are significant only for points $\gtrsim
3\sigma$ from the best fit values, where corrections to the likelihood surface
of $\xi_{0,2}$ will also become important. 

\subsection{Key degeneracies and error budget}
\label{errorbudget}
\begin{figure}
  \includegraphics[width=85mm]{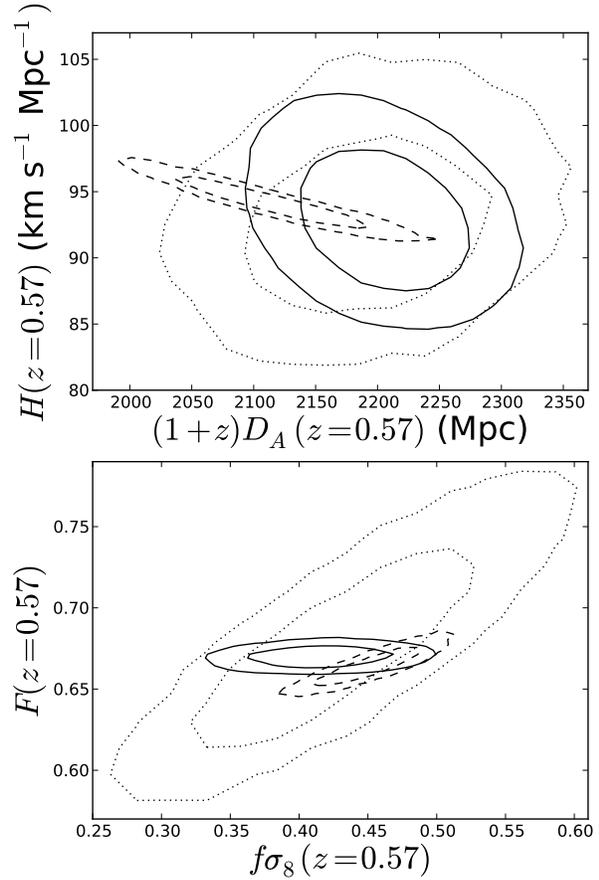}
  \caption{{\em Upper panel:} 68 and 95 per cent confidence regions for the
comoving angular diameter distance and expansion rate at $z=0.57$ from CMASS
anisotropic clustering constraints when
$f\sigma_8$ is varied over the WMAP7+GR flat $\Lambda$CDM prior (Model 3; solid) and
when $f\sigma_8$ is simultaneously fit (Model 4; dotted).  
{\em Lower panel:} 68 and 95 per cent confidence regions for $f\sigma_8(z=0.57)$ and
Alcock-Paczynski parameter $F(z=0.57)$ inferred from CMASS anisotropic clustering
(Model 4; dotted). These two parameters are partially degenerate, and their differing
scale-dependence allows us to constrain each separately.  The solid contour
shows the constraint when a WMAP7+GR flat $\Lambda$CDM prior is used on $F$.  
In both panels we show for comparison the predictions from
WMAP7 when a standard GR, flat $\Lambda$CDM cosmology is assumed (dashed).}  
  \label{fig:twodhist}
\end{figure}

In this section we examine the main sources of uncertainty in our measurements
of $f\sigma_8(z_{\rm eff})$ and $F(z_{\rm eff})$.  Both of these
parameters affect the amplitude of the quadrupole, and so are partially
degenerate;
Figure \ref{fig:twodhist} shows that our measurements are sufficiently sensitive to
distinguish the two through their differing scale-dependence.   We will
consider in turn the uncertainty due to
the nuisance parameter $\sigma^2_{\rm FoG}$ describing the intrahalo velocities 
of satellite galaxies (the ``finger-of-god effect''), uncertainty in the
underlying linear matter power spectrum,
the redshift-distance relation (in the case of peculiar
velocities), and the peculiar velocity field (in the case of the AP effect).

We first assess the impact of non-linearity in the covariance matrix on the
error budget.  Taking the number of galaxies in the present analysis, assuming 
 $\bar{n} = 3 \times 10^{-4}$ (h$^{-1}$ Mpc)$^{-3}$, and using linear theory
to evaluate the Fisher matrix \citep[as in][]{ReiWhi11}, we expect 
an uncertainty on $f\sigma_8$ of $\approx 0.021$ when
only $b\sigma_8$ and $f\sigma_8$ are freely varied. 
Using our mock covariance matrix, we find an uncertainty of 0.029 on $f\sigma_8$ 
at fixed $\sigma^2_{\rm FoG} = 21$ Mpc$^2$ and for $P_{\rm lin}(k)$, $D_V$, and
$F$ all fixed at their values in the mock catalog cosmology.  This $\sim 40$ per cent
increase is primarily due to the nonlinear/window function corrections 
to the covariance matrix highlighted in Figure \ref{fig:LPTvslinear}.
If we instead use the mock based covariance matrix to fit for both $D_V$ and
$F$, fixing $\sigma^2_{\rm FoG}$, $P_{\rm lin}(k)$, and $f\sigma_8$, we find
$\sigma_{F}=0.019$.
These results are summarized in Table \ref{table:uncertainties}.
\begin{table}
\begin{center}
\begin{tabular}{lllll}
Marginalized parameters & $\sigma_{b\sigma_8}$ & $\sigma_{f\sigma_8}$ &
$\sigma_{F}$ & Model \\
\hline
Fisher & 0.019 & 0.021 & - & - \\
Fisher, $\sigma^2_{\rm FoG}$ & 0.024 & 0.042 & - & -\\
\hline
- & 0.023 & 0.029 & - & -\\
$\sigma^2_{\rm FoG}$ & 0.025 & 0.033 & - & -\\
$\sigma^2_{\rm FoG}$, $P(k)$ & 0.033 & 0.033 & - & - \\
$\sigma^2_{\rm FoG}$, $P(k)$, geometry & 0.033 & 0.034 & - & 2\\
\hline
- & 0.037 & - & 0.019 & -\\
$\sigma^2_{\rm FoG}$ & 0.038 & - & 0.022 & - \\
$\sigma^2_{\rm FoG}$, $P(k)$ & 0.046 & - & 0.022 & - \\  
$\sigma^2_{\rm FoG}$, $P(k)$, growth & 0.046 & - & 0.026 & 3 \\
\hline
$\sigma^2_{\rm FoG}$ & 0.047 & 0.069 & 0.042 & - \\
$\sigma^2_{\rm FoG}$, $P(k)$ & 0.050 & 0.069 & 0.042 & 4 \\
\end{tabular}
\end{center}
\caption{We examine how uncertainty in various quantities entering our
analysis impacts the 68 per cent confidence level intervals on parameters
$b\sigma_8$, $f\sigma_8$, and $F$.  For comparison, the first two rows show
the predictions from a simple linear theory Fisher matrix analysis
\citep[as in][]{ReiWhi11} with
$\bar{n}=3\times 10^{-4}\,(h^{-1}{\rm Mpc})^{-3}$,
the number of galaxies in the present analysis, and ignoring all window
function effects.  When $\sigma^2_{\rm FoG}$ is marginalized over for the
measurements (but {\em not} for the Fisher matrix analysis), we maintain
a hard prior $0<\sigma^2_{\rm FoG}<40\,{\rm Mpc}^{2}$.
Uncertainty in the underlying linear $P(k)$ is derived from WMAP7 data,
under the assumptions of Gaussian, adiabatic, power law initial conditions
with $N_{\rm eff}=3.04$ massless neutrino species.
Uncertainty in geometry [$D_V(z_{\rm eff})$ and $F(z_{\rm eff})$] or growth
[$f\sigma_8$] is taken to be the uncertainty on these quantities derived from
WMAP7 in a $\Lambda$CDM cosmology.  The last column highlights the cases
corresponding to Models 2 though 4.}
\label{table:uncertainties}
\end{table}

\subsubsection{Degeneracy with $\sigma^2_{\rm FoG}$}
At fixed $P(k)$ and geometric parameters, the Fisher matrix analysis indicates
a factor of 2 increase in the $f\sigma_8$ error, to $0.042$, when
$\sigma^2_{\rm FoG}$ is marginalized over without any prior, compared to when
it is fixed at $\sigma^2_{\rm FoG}=21\,{\rm Mpc}^2$.
The marginalized error on $\sigma^2_{\rm FoG}$ in the former case is
$14\,{\rm Mpc}^2$.
Therefore, the hard prior $0<\sigma^2_{\rm FoG}<40\,{\rm Mpc}^2$ substantially
reduces this source of uncertainty.
Table \ref{table:uncertainties} indicates an increase of only 15 per cent in
the error on $f\sigma_8$, to 0.033, when we marginalize over
$\sigma^2_{\rm FoG}$ within our hard prior.
Similarly, marginalizing over $\sigma^2_{\rm FoG}$ increases the error on
$F$ from 0.019 to 0.022.
Therefore, further reduction in the uncertainty on $\sigma^2_{\rm FoG}$ with
more detailed modeling of the small-scale clustering would only allow a slight
reduction in the errors.
However, since our fits indicate a slight preference for
$\sigma^2_{FOG}=40\,{\rm Mpc}^2$ compared to the fiducial $21\,{\rm Mpc}^2$,
in future work we will revisit our choice of prior after a re-analysis of
small-scale CMASS clustering with a larger data set.

\subsubsection{Uncertainty in the underlying linear matter $P(k)$}

Of the three parameters defining the shape of the underlying linear matter
$P(k)$ in our analysis, the uncertainty in $\omega_c =\Omega_c h^2$ is
the largest, particularly in the context of the constraining power of the
CMASS measurements.  For this comparison,
we therefore hold $\Omega_b h^2 = 0.02258$ and $n_s = 0.963$ fixed, and examine how the
central values and uncertainties in  $b\sigma_8$, $f\sigma_8$ or $F$ change as a
function of $\omega_c$, with all other parameters held fixed.  Empirically,
the uncertainties on these parameters do not depend much on $\omega_c$.  
Therefore, we can estimate the impact on the uncertainty in $\omega_c$
 through the dependence of the central value of parameter $p$ on $\omega_c$,
 in units of the uncertainty on those quantities:
\begin{equation}
s = \frac{\Delta p/\sigma_p}{\Delta \omega_c/\sigma_{\omega_c}},
\end{equation}
where $\sigma_p$ is the uncertainty at fixed $\omega_c$.
For Gaussian probability distribution functions in 
$P(p|\omega_c)$ and $P(\omega_c)$, the uncertainty on $p$
when marginalized over $\omega_c$ is increased by $\sqrt{1+s^2}$.  We find that 
$s\lesssim 0.1$ for $f\sigma_8$ and $s = 0.13$ for $F$.  Therefore, the current
uncertainty in the underlying linear matter $P(k)$ is negligible for the purpose
of deriving constraints from the anisotropic clustering of CMASS galaxies.  This
justifies the use of a fixed power spectrum shape in the WiggleZ analyses of
anisotropic clustering \citep{Bla11b,Bla11c}.  However, the
uncertainty in $P(k)$ does increase the uncertainty in $b\sigma_8$, so
applications such as galaxy-galaxy lensing/galaxy clustering combinations
\citep[e.g.,][]{reyes/etal:2010} should marginalize over this additional uncertainty.

\subsubsection{Uncertainty in $\Lambda$CDM geometry}

Table \ref{table:uncertainties} indicates that marginalizing over the
uncertainty in the geometric quantities $(1+z_{\rm eff}) D_A(z_{\rm eff})$
and $H(z_{\rm eff})$ [or equivalently, $D_V(z_{\rm eff})$ and $F(z_{\rm eff})$]
contributes negligibly to the uncertainty in $f\sigma_8$ for WMAP7
uncertainties when a $\Lambda$CDM redshift-distance relation is assumed.

\subsubsection{Uncertainty in $\Lambda$CDM growth of structure}

In the $\Lambda$CDM model, WMAP7 constrains $\Omega_m$ to $\sim 11$ per cent;
this translates into a relatively large uncertainty in the predicted growth
rate of structure ($f\sigma_8$) of 5.5 per cent.
Marginalizing over this uncertainty increases our uncertainty on $F$ from
0.022 to 0.026.
Weak lensing \citep[e.g.,][]{Mun08}
and/or cluster abundances \citep[e.g.,][]{Vik09,rozo/etal:2010} 
reduce the $\Lambda$CDM uncertainty on a combination of $\sigma_8$ and
$\Omega_m$ similar to the GR prediction for peculiar velocities,
$\sigma_8 \Omega_m^{0.55}$, and thus could potentially be used to reduce
the uncertainty on $f\sigma_8$ in the $\Lambda$CDM growth scenario.
In the present work, we do not explore other dataset combinations besides
WMAP7 and CMASS.

\subsection{Comparison with previous measurements at $z \approx 0.6$}

The WiggleZ survey has recently analyzed the anisotropic clustering of bright
emission line galaxies over a broad redshift range \citep{Bla11b,Bla11c}.
Their growth rate constraints assumed a fixed underlying linear matter power
spectrum and redshift-distance relation, for which our error is 0.033.
With a factor $\sim 4$ fewer galaxies, they achieve comparable precision and
good agreement with our central value in their $z=0.6$ bin: 
$f\sigma_8(z=0.6) = 0.43 \pm 0.04$.  In their analysis they include modes up to
$k_{\rm max} = 0.3\,h\,{\rm Mpc}^{-1}$ and marginalize over a Lorentzian model
to account for small scale nonlinearities.
We find that the quoted WiggleZ error is in good agreement with our Fisher
matrix prediction if we assume
$\bar{n}=2.4\times 10^{-4}\,(h^{-1}{\rm Mpc})^{-3}$, $b=1.1$,
$N_{\rm gal}=60227$, a Gaussian damping
$\sigma_{\rm Gauss}=300\,{\rm km}\,{\rm s}^{-1}$ and
$s_{\rm min} = 1.15\pi\,k_{\rm max}^{-1}=12\,h^{-1}$Mpc,
i.e., a factor of two smaller scale than we have adopted for our analysis. 
Exploration using the Fisher matrix suggests that the difference in number
densities between CMASS and WiggleZ has a negligible impact on the
uncertainties, while the lower bias of their sample implies a 15 per cent
(10 per cent) improvement at $s_{\rm min}=25\,(12)\,h^{-1}$ Mpc.
By far the dominant difference arises because they are fitting out to
$k_{\rm max}=0.3\,h\,{\rm Mpc}^{-1}$;
their error would increase by a factor of 2 if they adopted our minimum scale.
Another issue is that our mock covariance matrix, which accounts for both
nonlinear growth of structure and our complex survey geometry, yield errors
40 per cent larger than our naive Fisher matrix analysis would predict, with
the difference between the linear and mock covariance matrices increasing on
small scales (for $\xi_2$).
While it is not clear how this difference scales with galaxy bias or depends on
survey geometry, the WiggleZ use of linear theory covariance matrices down to
much smaller scales could cause their uncertainty to be underestimated.
 
Another recent WiggleZ analysis \citep{Bla11c} simultaneously fits $F =
0.68 \pm 0.06$ and $f\sigma_8 = 0.37 \pm 0.08$ at $z = 0.6$ for a fixed
underlying $P(k)$, and makes use of modes with $k<0.2\,h\,{\rm Mpc}^{-1}$,
which can be compared with the last two lines of Table 
\ref{table:uncertainties} and Model 4 in Table \ref{tab:bf4b4}.
As in the previous discussion, our best fit values for these parameters
are consistent, and using clustering information on small scales
(i.e., larger $k_{\rm max}$) permits a factor of $\sim 1.4$ tighter constraint
on $F$ at a fixed number of galaxy spectra compared with our clustering
analysis of CMASS galaxies.  
However, the higher bias of CMASS galaxies permits relatively tighter
constraints on $F$ than $f\sigma_8$ as compared with WiggleZ.

The WiggleZ and BOSS CMASS surveys target very different galaxy types, which will have
different nonlinear properties and modeling uncertainties.  Without a detailed
study it is impossible to understand the robustness of various assumptions for
the nonlinear distortions, particularly  when multiple sources of nonlinearity
contribute. These can potentially cancel when examining clustering, meaning
that goodness-of-fit must be used carefully when assessing robustness and
potential biases in a model fit.
In this work we have adopted a conservative approach based on the best model
available that has been compared with a large volume of $N$-body mock
galaxy catalogs, and only fit over those scales where we are confident that the
signal is dominated by the quasilinear velocity field of interest, and where the
impact of small-scale random motions can be simply modeled and marginalized
over.

\section{Discussion}
\label{sec:discussion}
\begin{figure}
\begin{center}
\resizebox{3.2in}{!}{\includegraphics{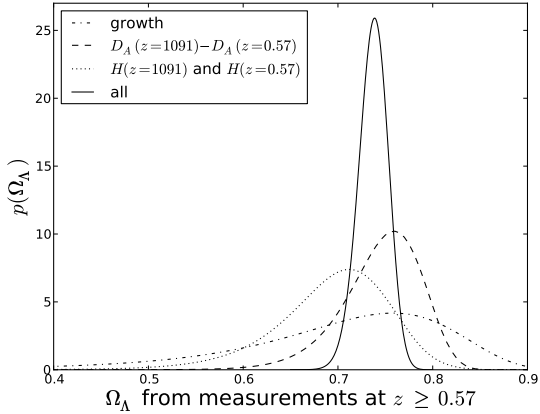}}
\end{center}
\caption{Joint constraints on $\Omega_{\Lambda}$ from the
combination of CMB distance and normalization priors and our measurements of 
$D_A$, $H$, and $f\sigma_8$.  These constraints are independent of the growth
and expansion history at redshifts lower than our galaxy sample, but assume a
flat $\Lambda$CDM cosmology between the CMASS sample and the CMB, as well as the
other model assumptions detailed in Section \ref{cosmomodel}.}
\label{fig:ol}
\end{figure}
We have analyzed the anisotropic clustering of BOSS DR9 CMASS galaxies with an accurate
analytic model for the monopole and quadrupole correlation functions, which we
have validated using a large volume of $N$-body based mock galaxy catalogs.  
The combination of the BAO standard ruler and the Alcock-Paczynski effect allows
us to separately constrain the comoving angular diameter distance and Hubble
expansion rate at the effective redshift of our sample, $z=0.57$, while redshift
space distortions allow us to constrain the amplitude of the peculiar velocity
field, a direct measurement of the growth {\em rate} of structure, ${\rm d}
\sigma_8/{\rm d} \ln a$.  Table \ref{tab:bf4b4}
summarizes our constraints under several assumptions.
Figure \ref{fig:twodhist} shows the degeneracy between the Alcock-Paczynski
parameter $F$ and the growth rate of structure in our measurements; this
explains why our constraints improve considerably if we make further model 
assumptions about the geometry or growth.  In the
most general case where all three parameters vary independently, 
we find ${\rm d}\sigma_8/{\rm d} \ln a = 0.43^{+0.069}_{-0.063}$, 
$(1+z_{\rm eff}) D_A(z_{\rm eff}) = 2190 \pm 61$ Mpc and 
$H(z_{\rm eff}) = 92.4^{+4.5}_{-4.0}$ km s$^{-1}$ Mpc$^{-1}$.

To illustrate the
cosmological constraining power of our measurements, we summarize our results as
three distinct tests of a minimal $\Lambda$CDM cosmology 
(see Section \ref{cosmomodel}) that link the observed CMB anisotropies at $z
\approx 1091$ and CMASS galaxy fluctuations at $z\approx 0.57$,
independent of the expansion history and growth of structure in the universe at
redshifts below our sample ($z \lesssim 0.57$).  We use a flat $\Lambda$CDM model to
illustrate our constraining power on the behavior of the universe at $z > 0.57$,
and explore more general cosmological models in our companion paper,
\citet{Samushia:prep}.\\

{\em Is the shape of the power spectrum of matter density fluctuations
inferred from the CMB consistent with the one inferred from galaxy
fluctuations after a factor of $\sim 4\times 10^5$ amplification?}\\

If the dominant component of the energy density is ``cold'', then in the
linear regime perturbation growth in General Relativity is scale-independent.
Modulo our corrections for non-linear $\Lambda$CDM evolution and galaxy biasing
($\sim 10$ per cent on the scales we analyse),
the shapes of the linear matter power spectrum inferred from the CMB and
galaxy clustering are consistent: our best fit $\Lambda$CDM model gives
$\chi^2 = 39.3$ for 44 degrees of freedom.  
We quantify this statement further using our fit to the location of the
broad turnover in $P(k)$, which indicates the horizon size at matter-radiation
equality and thus the physical matter density, assuming the radiation density
is known.
Combining our constraint ($\Omega_m h^2 = 0.142 \pm 0.011$) with those in the
literature for the SDSS-II LRG sample and WiggleZ yields $\Omega_m h^2 = 0.137
\pm 0.0064$, where the error neglects the expected small but non-zero covariance
between the galaxy samples, and also fixes $\Omega_b h^2$ and $n_s$ to best fit
CMB values.  The CMB constraint is only slightly more precise than the combined
galaxy measurement ($\Omega_m h^2 = 0.1334
\pm 0.0056$), and the two are in excellent agreement. 
With this test passed, in the rest of our analysis we make use of the full matter power 
spectrum (rather than just the BAO feature) as a standard ruler in galaxy clustering
measurements.  See \citet{Sanchez:2012} for an exploration of other cosmological
models using the CMASS monopole correlation function shape.  In particular, the
good agreement between constraints on the linear matter power spectrum from the
CMB and CMASS galaxy clustering limit the allowed contribution from species
such as massive neutrinos which induce scale-dependent growth.\\

{\em Do our constraints on the geometry of the universe require dark energy at
$z \gtrsim 0.57$?} \\

Distance constraints from both the CMB and the BAO feature are determined
relative to the sound horizon at $z_{\star} \approx 1091$, the redshift of
decoupling, and at $z_{\rm drag} \approx 1020$, the redshift when baryons
were released from the Compton drag of the photons\footnote{We follow the WMAP7
analysis and adopt the fitting formulae for $z_{\star}$ and $z_{\rm drag}$ from 
\citet{HuSug96} and \citet{EisHu98}, respectively.}.  
Within the constraints of our cosmological model assumptions (see
Section \ref{cosmomodel}), $r_s(z_{\star})$ and $r_s(z_{\rm drag})$ depend only on $\Omega_m
h^2$ and $\Omega_b h^2$.  We project the observed {\em difference} in comoving
angular diameter distance
$\Delta D_A = (1+1091) D_A(z=1091)-(1+0.57) D_A(z=0.57)$,  
\begin{equation}
\Delta D_A = \int_{0.57}^{1091} \frac{dz}{H(z)}
\end{equation}
onto the parameter $\Omega_\Lambda$ in a flat $\Lambda$CDM cosmology 
and marginalize over the other parameters.  The result is shown by the dashed
line in Figure \ref{fig:ol}: $\Omega_\Lambda = 0.76 \pm 0.04$.  We quote the
maximum likelihood and 68 per cent confidence region around it unless otherwise noted.
  We have purposely chosen
a variable that is independent of the expansion history at $z < 0.57$.
Similarly, the CMB constraint 
on $\Omega_m h^2$ with our measurement of $H(z=0.57)$ requires $\Omega_\Lambda =
0.71_{-0.05}^{+0.06}$ between $z=0.57$ and $z=1091$ (dotted curve in Figure
\ref{fig:ol}).\\

{\em Does the observed growth rate of structure at $z = 0.57$ require dark
energy?}\\

Because WMAP places such a tight constraint on the
amplitude of curvature perturbations deep in the matter-dominated epoch 
at $k = 0.027$ Mpc$^{-1}$ (1.8 per cent), we can
translate our measurement of $f\sigma_8$ into a constraint on ${\rm d}D/{\rm d}
\ln a$, where $D$ is the usual growth function that depends only on $\Omega_m$
or $\Omega_\Lambda$ in a flat $\Lambda$CDM cosmological model:
\begin{equation}
\label{growth}
  D(z) = \frac{5\Omega_m}{2}\frac{H(z)}{H_0}
         \int^{1/(1+z)} \frac{da\ H_0^3}{[aH(a)]^3}.
\end{equation}
To do so, we require the additional cosmological model assumptions listed in
Section \ref{cosmomodel}.  We marginalize over the WMAP7 uncertainties in the 
parameters that convert curvature perturbations to the integrated amplitude of matter
perturbations on scales of $R = 8/0.7 = 11.4$ Mpc, $\sigma_R$, namely 
$\Omega_m h^2$, $\Omega_b h^2$, and $n_s$.  
Section \ref{sigma8def} details the relation between $\sigma_R$ and $\sigma_8$,
which can be regarded as equal for our purposes.
This comparison between the fluctuation amplitude at $z=1091$ and $z=0.57$
requires $\Omega_{\Lambda}$ within $[0.59,0.81]$ (central 68 per cent confidence), with a maximum
likelihood at 0.76.  The distribution is shown as the dot-dashed curve in Figure
\ref{fig:ol}.\\

In combination with the CMB, both our geometric and growth rate constraints 
require a value of $\Omega_\Lambda$ at
$0.57 < z < 1091$ that is consistent with the concordance model expectation.
Combining all three constraints with the WMAP7 distance and normalization
priors, and marginalizing over $\Omega_{b} h^2$ and
$\Omega_m h^2$, we find $\Omega_\Lambda = 0.74_{-0.015}^{0.016}$ (solid curve in
Figure \ref{fig:ol}).\\

Our analysis required further assumptions compared with a BAO-only
analysis \citep{Aardvark} that constrains $D_V(z_{\rm eff})/r_s(z_{\rm drag})$. 
However, we were able to perform tests of the scale-independence and
growth rate of cosmic structure between recombination and $z \approx 0.57$, 
as well as to break the degeneracy between $(1+z_{\rm eff}) D_A$ and $H$ using the
Alcock-Paczynski test.  So far, our measurements do not unveil any deviations
from the minimal $\Lambda$CDM model we have examined.  
Under the assumption
of a $\Lambda$CDM cosmology extending to $z=0$, 
the BAO feature adds the most constraining power
to WMAP7 on $\Lambda$CDM parameters; we find $\Omega_m h^2 = 0.136 \pm 0.0035$,
$\Omega_m = 0.283 \pm 0.017$, and $H_0 = 69.3 \pm 1.5$ km s$^{-1}$ Mpc$^{-1}$ in
this model.  We anticipate statistical improvements on these results with the
completed BOSS galaxy dataset covering a footprint three times larger,  
as well as developments in the theoretical modeling that will allow tighter
cosmological constraints.

\section*{Acknowledgements}
Funding for SDSS-III has been provided by the Alfred P. Sloan
Foundation, the Participating Institutions, the National Science
Foundation, and the U.S. Department of Energy Office of Science.
The SDSS-III web site is http://www.sdss3.org/.

\noindent SDSS-III is managed by the Astrophysical Research Consortium for the
Participating Institutions of the SDSS-III Collaboration including the
University of Arizona,
the Brazilian Participation Group,
Brookhaven National Laboratory,
University of Cambridge,
Carnegie Mellon University,
University of Florida,
the French Participation Group,
the German Participation Group,
Harvard University,
the Instituto de Astrofisica de Canarias,
the Michigan State/Notre Dame/JINA Participation Group,
Johns Hopkins University,
Lawrence Berkeley National Laboratory,
Max Planck Institute for Astrophysics,
Max Planck Institute for Extraterrestrial Physics,
New Mexico State University,
New York University,
Ohio State University,
Pennsylvania State University,
University of Portsmouth,
Princeton University,
the Spanish Participation Group,
University of Tokyo,
University of Utah,
Vanderbilt University,
University of Virginia,
University of Washington,
and Yale University.

\noindent The simulations used in this paper were analysed at the
National Energy Research Scientific Computing Center, the Shared Research
Computing Services Pilot of the University of California and the
Laboratory Research Computing project at Lawrence Berkeley National Laboratory.

\noindent BAR gratefully acknowledges support provided by NASA through Hubble Fellowship grant 51280 awarded by the Space Telescope Science Institute, which is operated by the Association of Universities for Research in Astronomy, Inc., for NASA, under contract NAS 5-26555.
MW is supported by the NSF and NASA.
LS and WP thank the European Research Council for support. WP also acknowledges support from the UK Science and Technology Facilities Council.
MECS was supported by the NSF under Award No. AST-0901965.

\appendix
\section{Observational Uncertainties}
\subsection{Uncertainties in radial distribution}
\label{app:nz}

To compute overdensities of the galaxy field, defined as 
\begin{equation}
  \label{eq:overdensities}
  \delta(z,\hat{\Omega}) = \frac{\rho_0(z,\hat{\Omega}) -
  \rho(z,\hat{\Omega})}{\rho_0(z,\hat{\Omega})},
\end{equation}
\noindent
we must know the unperturbed density field $\rho_0(z,\hat{\Omega})$. While
the angular selection function of the survey is usually well known, the radial
distribution is not easy to model accurately. Usually, the unperturbed radial
distribution of galaxies is modeled from the distribution of observed redshifts
by either shuffling them or splining with a smooth curve. Different ways of
constructing a random catalog will result in different estimates of correlation
function \citep[see, e.g.,][]{SamPerRac11,Ross:2012}. 

To estimate the magnitude of this effect on the measurements of the moments of
correlation function we take 600 mock catalogs of CMASS sample and apply
simplified version of our analysis to the measurements produced using different
ways of reconstructing $n(z)$. We reconstruct $n(z)$ first by shuffling ``observed''
redshifts in mock catalogs and then by splining that distribution with 10, 20
and 30 node cubic spline fits. 

We find that the ways of reconstructing $n(z)$ for random catalogs do not affect
the RSD measurements significantly. In our current analysis we are using the
``shuffled'' catalogs since they introduce the least bias in the
measurements of $\xi_\ell(r)$ \citep{Ross:2012}.

\subsection{Effects of Close Pairs, redshift errors and redshift failures}
\label{app:closepairs}
We are unable to obtain redshifts for $\sim$ 5 per cent of the galaxies due to
fiber collisions -- no two fibers on any given observation can be placed closer
than $62''$.  At $z\simeq 0.5$ this $62''$ exclusion corresponds to
$0.4\,h^{-1}$Mpc.
Redshifts for some of the collided galaxies can be reclaimed in regions where
plates overlap, but the remaining exclusion must be accounted for.  We account
for fiber-collided galaxies by assigning its weight to its nearest neighbor on
the sky.  
Tests on mock catalogs presented in \citep{Guo11} indicate that the nearest 
neighbor correction adopted in this work is accurate to better than 1 per cent for 
both $\xi_{0}$ and $\xi_{2}$ at the scales used in our analysis.

Redshift failures are discussed in detail in \citet{Ross:2012}; we also correct
for them with a nearest-neighbor upweighting scheme.  Redshift measurement
errors smooth the apparent galaxy density field,
in the same fashion as described with our nuisance parameter $\sigma^2_{\rm FoG}$.
The median redshift error for our sample is 42 km s$^{-1}$, which translates
into an additive contribution to $\sigma^2_{\rm FoG} < 1$ Mpc$^2$. 

\section{Accuracy and implementation of the theoretical model}
\label{cosmoxi2d}
\subsection{Wide Angle Effects}
Equations \ref{eq:kaiser} and \ref{streamingeqn} assume that a
``plane-parallel'' approximation, which states that the sample is far enough
from the observer so that all line-of-sights are parallel to each other, is
accurate enough.
This approximation will fail at some scale for wide surveys
\citep[see e.g.][]{PapSza08}.
We approximate the scale dependent magnitude of wide-angle effects for
our sample in a similar manner to \citet{SamPerRac11}.
The effective redshift of our galaxy sample is $z=0.57$, which in the
best-fit WMAP7 cosmology corresponds to a comoving distance of approximately
$1500\Mpcoh$.
The largest scale that we consider in this analysis is $160\Mpcoh$,
which corresponds to an opening angle of about 3 degrees.
We estimate the wide-angle corrections to be at most 10 per cent of the
statistical errors on the largest scales and do not try to correct for this
effect in current analysis.

\subsection{{\it cosmoxi2d} code implementation}

\citet{ReiWhi11} demonstrated that Eqn.~\ref{streamingeqn} provides an accurate
description of $\xi^{s}_g(r_{\sigma},r_{\pi})$ when real space clustering and
velocity statistics inside the integrand are measured directly from $N$-body
simulations.  Moreover, in the regime where $b \approx 2$ (appropriate for CMASS
galaxies), the real space clustering and velocity statistics can be computed
analytically with sufficient precision to predict
$\xi^{s}_g(r_{\sigma},r_{\pi})$.  In detail, this ``sweet spot'' in the
precision of the model arises because the functions $v_{12}(r)$ and
$\sigma_{12,\parallel/\perp}^2(r)$ entering  Eqn.~\ref{streamingeqn} were 
evaluated as a function of cosmological parameters in standard perturbation theory 
accounting for only the linear bias of the tracer.  At the redshift of interest, 
the second order bias $b_{2L}$ crosses zero near $1+b_{1L} =
2$, and the calculation of these functions neglecting $b_{2L}$ is sufficient for our purposes.  
We therefore caution
against the use of our code for tracers with bias substantially different from
2.  Note that $\xi^{r}_g(r)$ is evaluated
in LPT \citep{Mat08b} and includes the contribution to
the real space clustering from second-order Lagrangian bias\footnote{We
relate the second order Lagrangian bias, $b_{2L}$, to the first order bias,
$b_{1L}$, through the peak background split; these parameters are not
varied independently.}.\\

Standard perturbation theory is known to have inaccuracies in
describing the BAO feature, and we found that on BAO scales errors in our
perturbation theory calculation of pairwise velocity statistics caused
inaccuracies in the prediction for $\xi^{s}_g(r_{\sigma},r_{\pi})$.  However, on
scales $s \gtrsim 70 \; h^{-1}$  Mpc, the redshift space version of 
LPT \citep{Mat08b} is very
accurate \citep[see figure 2 of][]{ReiWhi11}.  We therefore interpolate between
our evaluation of Eqn.~\ref{streamingeqn} at smaller scales and LPT
on large scales, with the transition fixed at 100 Mpc.

Finally, we model the effect of galaxy intrahalo velocities (traditional
``fingers-of-god'') by convolving our model $\xi(r_p, r_{\pi})$ with an
additional Gaussian velocity dispersion of variance $\sigma^2_{\rm FoG}$.  
Using the Gaussian form allows  
$\sigma^2_{\rm FoG}$ to be included directly in the Gaussian in 
Eqn.~\ref{streamingeqn} for faster evaluation of our model as we explore
cosmological parameter space.

We have developed a code to numerically evaluate Eqn.~\ref{streamingeqn}
as well as all the relevant perturbation theory integrals as a function
of an input linear matter power spectrum and nuisance parameters $b$ and
$\sigma^2_{\rm FoG}$.
The Alcock-Paczynski effect is easily incorporated using Eqn.~\ref{eq:APeqn}
before computing Legendre polynomial moments from
$\xi^{s}_g(r_{\sigma},r_{\pi})$.
The internal units of {\bf cosmoxi2d} are Mpc, in which the underlying
linear matter power spectrum is most tightly constrained.
The code is publicly available\footnote{http://mwhite.berkeley.edu/CosmoXi2D}.

\subsection{Model Accuracy}
\label{sec:modelacc}
Because we have such a large volume of simulations, we can use the difference
between the theoretical model at the known cosmological parameters of the 
$N$-body simulation and the measured correlation function from the mock galaxy 
catalogs to quantify our theoretical systematic error; we find 
$\Delta \chi^2 = 0.29$ at the best fit value of $\sigma^2_{\rm FoG}$.  We also 
compare the scale dependence of the model error with the 5 parameters we are 
fitting, and find $<0.25\sigma$ shifts compared to the {\em unmarginalized} 
uncertainties on all parameters (i.e., the uncertainty on each parameter 
if all the others were known perfectly).  Therefore we conclude that our 
systematic error is negligible in the context of this analysis.

Another concern is that the model becomes inaccurate rapidly on scales smaller
than our minimum fitting scale, $s_{\rm min}$.  When Alcock-Paczynski parameters vary,
 scales smaller than $s_{\rm min}$ contribute to the model.  However, we have verified
that for variations of $10$ per cent in $D_V$ and/or $20$ per cent in $F$ 
(i.e., much larger than the final uncertainties), the theoretical error induces 
$< 0.5\sigma$ shifts in all parameters compared with the unmarginalized uncertainties.

Finally, we point out that unlike in linear theory, our model depends on 
$b$, $f$, and $\sigma_8$ separately, rather than the only the combinations 
$b\sigma_8$ and $f\sigma_8$.  However, at the particular $b\sigma_8$ and $f\sigma_8$
 values of our sample, we find that changes in the predicted $\xi_{0,2}$ with $\sigma_8$
 can be absorbed by changes in the value of $b\sigma_8$.  Quantitatively, for a $\pm 10$ per cent
  change in away from our fiducial $\sigma_8(z_{\rm eff}) = 0.61$ and at fixed $P(k)$, AP parameters,
   and $\sigma^2_{\rm FoG} = 21$ Mpc$^2$, $b\sigma_8$ shifts by $\approx \pm 1.3 \sigma$ 
  with no measurable shift in the central value of $f\sigma_8$.
  
\subsection{Propagating uncertainties in the galaxy-halo mapping to $\sigma^2_{\rm FoG}$}
\label{galaxydetails}
\begin{figure}
\begin{center}
\resizebox{3.2in}{!}{\includegraphics{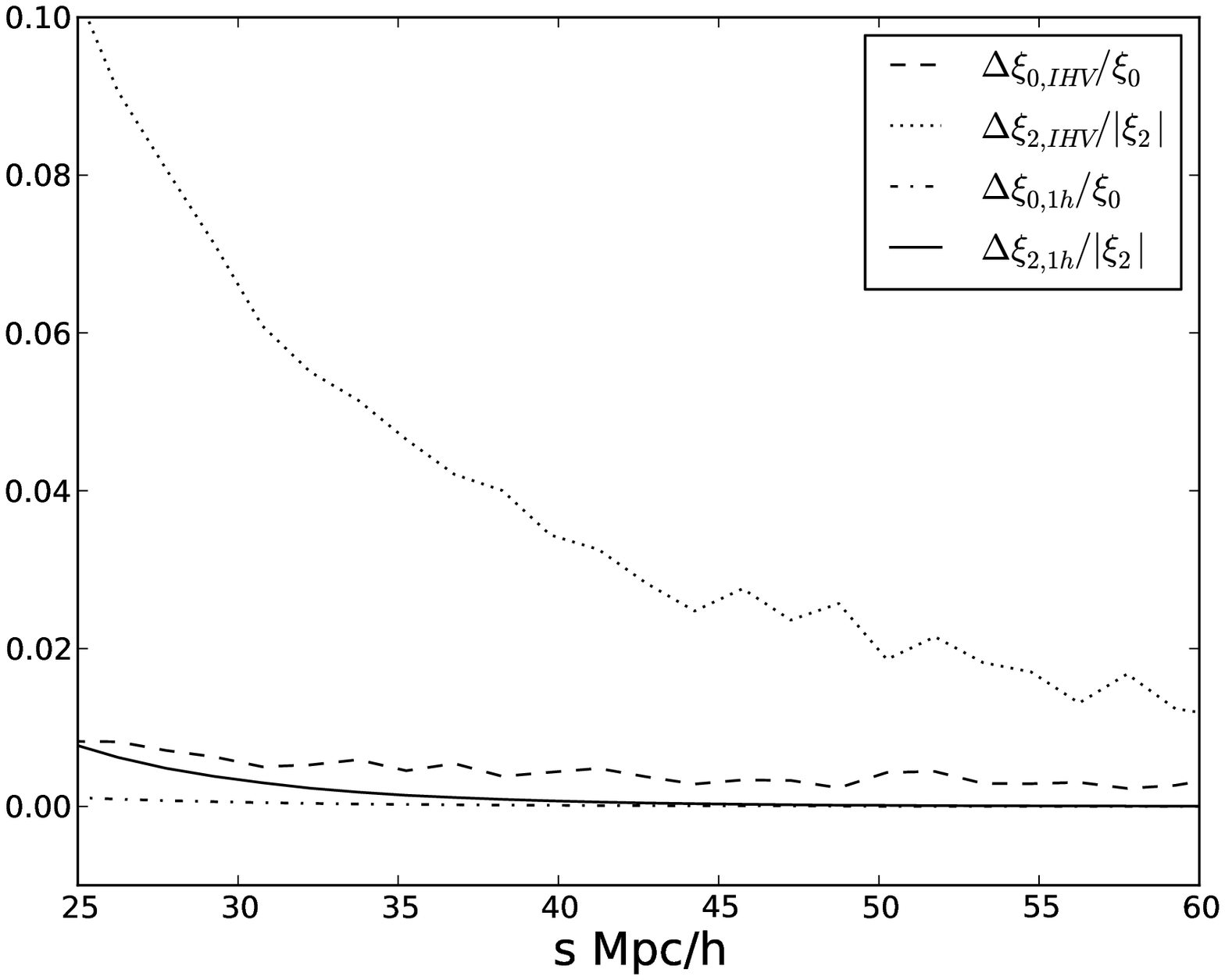}}
\end{center}
\caption{The dashed (dotted) lines show the fractional change in $\xi_0$
($\xi_2$) in our mock galaxy catalogs due to satellite galaxy intrahalo
velocities (IHV; also shown in Figure \ref{fig:theoryfits}).  We isolate the
contribution of pairs of galaxies occupying the same halo in our mock catalogs
(i.e., ``fingers-of-god''), shown as the dot-dashed (solid) curves for $\xi_0$
($\xi_2$).  On the scales of interest, the dominant effect of IHV is a net
diffusion of pairs from small scales (where $\xi$ is larger) to larger scales.}
\label{fig:1hvs2h}
\end{figure}

As shown in Figure \ref{fig:theoryfits}, our model with $\sigma^2_{\rm FoG} = 21 \;
{\rm Mpc}^2$ fits the $\xi_{0,2}(s)$ of our mock galaxy catalogs, based on the
best-fitting HOD in \citet{Whi11}. In this section we quantify how
uncertainties in both the theoretical modeling and data analysis cause
uncertainty in the expected value of $\sigma^2_{\rm FoG}$ for the CMASS sample.  We
address several aspects of this problem separately.
\begin{itemize}
\item {\em One-halo vs two-halo contributions to $\xi_{0,2}$ and fiber
collision corrections}: The formalism of the halo model distinguishes between
``one-halo'' and ``two-halo'' pairs depending on whether the two galaxies
occupy the same or different halos.  In Figure \ref{fig:1hvs2h} we show the
total change in $\xi_{0,2}$ due to satellite galaxy intrahalo velocities (IHV)
by the dashed (dotted) curves in our mock catalogs.  The dash-dot and solid
curves show the contributions to this change in $\xi_{0,2}$ from one-halo
pairs, which are localized to along the LOS with $r_{\sigma} \lesssim 1 \;
h^{-1} {\rm Mpc}$.  Because the one-halo pairs contribute to such a small
$d\mu$, they can be neglected on the scales used in our cosmological parameter
fits, $s \geq 25 \; h^{-1} {\rm Mpc}$.  This fact is important to establish since
our method for fiber collision corrections will correctly recover the
distribution of pair separations for pairs of galaxies with separations larger
than the fiber collision scale, but suppresses the contribution of pairs of
galaxies at the fiber collision scale ($r_{\sigma} \lesssim 0.5 \; h^{-1} {\rm
Mpc}$).
\item {\em Uncertainty in the HOD at fixed cosmology}: Uncertainties in the HOD
parameters will introduce an uncertainty in $\sigma^2_{\rm FoG}$.  While
$\sigma^2_{\rm FoG}$ is roughly proportional to the satellite fraction, it also
depends on the distribution of host halo mass -- increasing $\alpha$ and
$\kappa$ increases host halo mass at fixed satellite fraction, which increases
$\sigma^2_{\rm FoG}$.  We use MCMC chains from \cite{Whi11} to estimate our
uncertainty on $\sigma^2_{\rm FoG}$ at fixed cosmology to be 6 Mpc$^{2}$.
\item {\em Breakdown of the ``central'' galaxy assumption}: The analysis of
\cite{Ski11} suggests that the brightest galaxy in a halo is not always the
``central'' one.  We test the impact of relaxing our assumption that the
velocity of mock central galaxies have no intrahalo velocities by assigning
them the intrahalo velocity of a random dark matter particle halo member in our
simulations some fraction $p$ of the time.  When we have more than one galaxy
in a halo, we assume that the chance of not including the ``central'' galaxy in
our sample is lower, $\propto p^{N_{gal}}$.  For $p=0.3$, we find
$\sigma^2_{\rm FoG}$ at our fiducial HOD is increased by 9 Mpc$^2$.
\item {\em Variations in the halo mass function with cosmological parameters}:
A broad range of observations shows good agreement between the concordance
$\Lambda$CDM halo mass function and the multiplicity of galaxy groups and
clusters \citep[e.g., ][]{rozo/etal:2010,Allen:2011}, so this uncertainty is subdominant:
even if our fiducial HOD masses were scaled by a factor of 2, $\sigma^2_{\rm FoG}
\propto M^{2/3}$ would change by 11 Mpc$^2$.
\end{itemize}
Given the above considerations, we adopted a generous hard prior on $\sigma^2_{\rm FoG}$
between 0 and 40 Mpc$^2$.

\section{Markov Chain Monte Carlo (MCMC) Methods}
\label{sec:mcmcmethods}
We adopt a hybrid MCMC/importance resampling approach to explore the BOSS
likelihood surface with various priors imposed from the WMAP7 likelihood in
Models 1-4.  This approach is necessary in our case because our model evaluation
is slow, and we must marginalize over $b\sigma_8$ and $\sigma^2_{\rm FoG}$ at each
point in parameter space that we consider.  This section describes our
methods in each case.

\subsection{Importance resampling}
WMAP7 MCMC chains are publicly
available\footnote{http://lambda.gsfc.nasa.gov/product/map/dr4/parameters.cfm}.
These chains provide a fair sample of the WMAP7 likelihood surface.  Importance
resampling \citep{MCMC,lewisMCMC} allows us to compute how constraints on the model parameters
change given an additional constraint by evaluating the new likelihood at a
subsample of the original MCMC chain, multiplying the original weight of each
element by the new likelihood, and then recomputing confidence intervals.

\subsection{Model 1: WMAP7+CMASS $\Lambda$CDM}
For this model we marginalize only over two parameters, so it is feasible to
directly compute a marginalized CMASS likelihood:
\begin{equation}
P_{\rm CMASS}({\theta}_{\Lambda {\rm CDM}}) = \int db\sigma_8 \, d\sigma^2_{\rm FoG} \, e^{-\chi^2_{\rm CMASS}({\theta}_{\Lambda {\rm CDM}}, b\sigma_8, \sigma^2_{\rm FoG})/2}.
\end{equation}
We then use importance resampling of the WMAP7 chain parameters.

\subsection{Model 2 WMAP7+CMASS $\Lambda$CDM geometry, free growth; Model 3 WMAP7+CMASS $\Lambda$CDM growth, free geometry}
In Models 2 and 3, for each linear matter power spectrum, we must vary three or
four extra parameters describing the galaxy clustering observations.  We
therefore explore the likelihood surface at each point in the WMAP7 chain by
MCMC, and thereby sample the CMASS likelihood distribution $P(\theta_{\rm CMASS} |
\theta_{\Lambda {\rm CDM}})$.  In Model 2, $\theta_{\rm CMASS} = \{b\sigma_8,
\sigma^2_{\rm FoG}, f\sigma_8\}$, and in Model 3, $\theta_{\rm CMASS} = \{b\sigma_8,
\sigma^2_{\rm FoG}, D_V, F\}$.  MCMC chains at a fixed $\theta_{\Lambda {\rm
CDM}}$ sample parameter space proportional to $P(\theta_{\rm CMASS} | \theta_{\Lambda
{\rm CDM}})$, but in order to compute the marginalized likelihood of $f\sigma_8$
in Model 2 or $D_V, F$ in Model 3, we must  determine the relative
likelihood of the MCMC chains evaluated at different $ \theta_{\Lambda {\rm
CDM}}$.  Since
\begin{equation}
N e^{-\chi^2(\theta_{\rm CMASS}, \theta_{\Lambda {\rm CDM}})/2} = P(\theta_{\rm
CMASS}, \theta_{\Lambda {\rm CDM}}) = P(\theta_{\rm CMASS} | \theta_{\Lambda {\rm CDM}}) P(\theta_{\Lambda {\rm CDM}})
\end{equation}
where $N$ is an overall normalization, we can combine the $\chi^2$ computed at any
point with the probability density estimated by our MCMC to determine the
relative normalization of $P(\theta_{\Lambda {\rm CDM}})$.  In practice, we
first normalize each MCMC distribution so that $\int P(\theta_{\rm CMASS} |
\theta_{\Lambda {\rm CDM}}) = 1$, find $\theta^{\star}_{\rm CMASS}(\theta_{\Lambda
{\rm CDM}})$ with the minimum value of $\chi^2$ in each chain, and integrate
$P(\theta_{\rm CMASS} | \theta_{\Lambda {\rm CDM}})$ in a small, fixed size region of
parameter space around $\theta^{\star}_{\rm CMASS}$, which we call
$\tilde{p}(\theta^{\star}_{\rm CMASS}, d\theta_{\rm CMASS})$.
\footnote{A further subtlety is that to compute $\tilde{p}$, we restrict $\theta^{\star}$
to be sufficiently far from the hard prior boundary of $\sigma^2_{\rm FoG}$, so that
the volume element $d\theta_{\rm CMASS}$ does not intersect the hard boundary.  The
difference between $\chi^2(\theta^{\star})$ and the global $\chi^2$ minimum is
small, since $\sigma^2_{\rm FoG}$ is poorly constrained by our measurements.}
  The relative weight of each point in the WMAP-only MCMC chain is then 
  determined by the CMASS-only likelihood value and $\tilde{p}$:
\begin{equation}
w(\theta_{\Lambda {\rm CDM}}) = e^{-\chi^2(\theta^{\star}_{\rm CMASS} |
\theta_{\Lambda {\rm CDM}})/2} (\tilde{p}(\theta^{\star}_{\rm CMASS},
d\theta_{\rm CMASS}))^{-1}.
\end{equation}
We find that our constraints are the same if we neglect this volume weighting factor $\tilde{p}^{-1}$,
 indicating that the effective volume of parameter space allowed by the CMASS measurements
does not strongly depend on the underlying cosmological parameters
$\theta_{\Lambda {\rm CDM}}$ when exploring the region of this parameter space
allowed by WMAP7. 

\subsection{Model 4 WMAP7+CMASS, free growth, free geometry}
In this case, we use WMAP data to provide a prior on the shape of the linear
matter power spectrum, which is well approximated by a multivariate Gaussian in
the parameters $\Omega_c h^2$, $\Omega_b h^2$, $n_s$.  WMAP7 constraints on
these parameters primarily come from ratios of peak heights and the overall
shape, rather than the locations of the peaks (which are sensitive to the angular
diameter distance to the last scattering surface).  Therefore, marginalized
likelihood for these parameters is nearly independent of the adopted model for
the low redshift expansion history (i.e., whether $\Omega_k$, or $w$ are freely
varied).  Thus we are able to make use of the CMB information on the underlying
linear matter power spectrum that is independent of the model for the low
redshift expansion history, and does not contain information on the distance to
the last scattering surface.  By using the linear matter power spectrum as a
``standard ruler'', we are able to infer information about the geometric
parameters $D_V(z_{\rm eff})$ and $F$.  In this case we run an MCMC chain
with the following 8 parameters, adding an additional multivariate Gaussian
likelihood representing the CMB prior on the three linear matter power spectrum
parameters: $\{\Omega_c h^2, \Omega_b h^2, n_s, D_V(z_{\rm eff}), F,
b\sigma_8, f\sigma_8, \sigma^2_{\rm FoG}\}$.

\end{document}